\documentclass[review]{elsarticle}
\usepackage[utf8]{inputenc}
\usepackage{hyperref,amsmath,amssymb,amsbsy,bbold,color,graphics,graphicx,bigints,relsize,cancel,tcolorbox}
\usepackage{lineno,hyperref,appendix}

\modulolinenumbers[5]
\pdfoutput=1
\allowdisplaybreaks

\journal{Journal of \LaTeX\ Templates}

\begin{document}

\begin{frontmatter}

\title{Relativistic thermodynamics of perfect fluids}

\author{Sylvain D. Brechet}
\author{Marin C. A. Girard}
\address{Institute of Physics, Station 3, Ecole Polytechnique F\'ed\'erale de Lausanne\,-\,EPFL, CH-1015 Lausanne,\,Switzerland}

\begin{abstract}

\noindent The relativistic continuity equations for the extensive thermodynamic quantities are derived based on the divergence theorem in Minkowski space outlined by Stückelberg. This covariant approach leads to a relativistic formulation of the first and second laws of thermodynamics. The internal energy density and the pressure of a relativistic perfect fluid carry inertia, which leads to a relativistic coupling between heat and work. The relativistic continuity equation for the relativistic inertia is derived. The relativistic corrections in the Euler equation and in the continuity equations for the energy and momentum are identified. This relativistic theoretical framework allows a rigorous derivation of the relativistic transformation laws for the temperature, the pressure and the chemical potential based on the relativistic transformation laws for the energy density, the entropy density, the mass density and the number density.
	
\end{abstract}

\end{frontmatter}


\tableofcontents


\section{Introduction}

\noindent In his seminal paper entitled ``On the electrodynamics of moving bodies''~\cite{Einstein:1905} and published in the ``Annalen der Physik'' in 1905 during the ``annus mirabilis'', Albert Einstein laid the foundation of special relativity by rooting firmly his theory on the transformation laws for the electromagnetic fields derived independently in 1899 by Hendrik A. Lorentz~\cite{Lorentz:1899} and in 1900 by Henri Poincaré.~\cite{Poincare:1900} Shortly thereafter, an important effort was made to establish relativistic theories for all branches of physics, notably by Albert Einstein himself~\cite{Einstein:1907} in 1907. Max Planck~\cite{Planck:1908} derived in 1908 the first relativistic transformation law for the temperature,
\begin{equation}\label{Planck temperature frames}
T = \gamma^{-1}\,\tilde{T}  \qquad\textrm{(Einstein and Planck)}
\end{equation}
where $\tilde{T}$ is the temperature in the rest frame $\mathcal{\tilde{R}}$ and $T$ is the temperature in an inertial frame $\mathcal{R}$ and $\gamma > 0$ is the Lorentz factor. This transformation law was at the heart of the reigning paradigm of relativistic thermodynamics for over 50 years adopted notably by Louis de Broglie~\cite{Broglie:1948}, Olivier C. de Beauregard~\cite{Beauregard:1949}, Max von Laue~\cite{Laue:1952}, Richard C. Tolman~\cite{Tolman:1950} and Wolfgang Pauli~\cite{Pauli:1958}. Heinrich Ott called into question this paradigm.~\cite{Ott:1963} Taking into account the relativistic transformation law for the mass, i.e. $M = \gamma\,\tilde{M}$, where $\tilde{M}$ is the mass in the rest frame $\mathcal{\tilde{R}}$ and $M$ is the mass in an inertial frame $\mathcal{R}$, Ott showed in 1963 that the relativistic transformation law for the temperature becomes,~\cite{Ott:1963}
\begin{equation}\label{Ott temperature frames}
T = \gamma\,\tilde{T} \qquad\textrm{(Ott and Arzelies)}
\end{equation}
and Henri Arzelies reached independently the same conclusion in 1965.~\cite{Arzelies:1965} The difference between the relativistic transformation laws~\eqref{Planck temperature frames} and~\eqref{Ott temperature frames} for the temperature is referred to unfortunately as the Einstein and Ott controversy. A private letter sent by Einstein to von Laue in 1952 shows that Einstein's point of view clearly changed over time.~\cite{Liu:1992} The transformation law for the temperature derived in this letter is identical to the transformation law~\eqref{Ott temperature frames} derived later by Ott. So the controversy should be renamed adequately the Planck and Einstein controversy. A third relativistic transformation law of the temperature was postulated by Peter T. Landsberg~\cite{Landsberg:1967} and Nicolaas G. van Kampen~\cite{Kampen:1968} who argued independently in 1967 and 1968 that the temperature should be a Lorentz invariant quantity, 
\begin{equation}\label{Landsberg temperature frames}
T = \tilde{T} \qquad\textrm{(Landsberg and van Kampen)}
\end{equation}
All these authors agree on the frame independence of the entropy that should be a Lorentz invariant quantity as advocated initially by Albert Einstein~\cite{Einstein:1907} and Max Planck,~\cite{Planck:1908}
\begin{equation}\label{Planck entropy frames}
S = \tilde{S} \qquad\textrm{(Einstein, Planck et al.)}
\end{equation}
where $\tilde{S}$ is the entropy in the rest frame $\mathcal{\tilde{R}}$ and $S$ is the entropy in an inertial frame $\mathcal{R}$. Since the entropy is Lorentz invariant and the temperature is the intensive conjugate variable of the entropy, the relativistic transformation law for the internal energy has to be the same as the relativistic transformation law for the temperature. The relativistic transformation law for the pressure is another essential result for a relativistic theory of thermodynamics. Albert Einstein~\cite{Einstein:1907}, Max Planck~\cite{Planck:1908} and Arzelies~\cite{Arzelies:1965} reached the conclusion that the pressure is Lorentz invariant,
\begin{equation}\label{Einstein pressure frames}
p = \tilde{p} \qquad\textrm{(Einstein, Planck and Arzelies)}
\end{equation}
where $\tilde{p}$ is the pressure in the rest frame $\mathcal{\tilde{R}}$ and $p$ is the pressure in an inertial frame $\mathcal{R}$. These authors used the mechanical definition of the pressure as the ratio of the normal force and the surface area.~\cite{Agmon:1977} Sutcliffe derived another transformation law using the thermodynamic definition of the pressure as the intensive conjugate variable of the volume in 1965,~\cite{Sutcliffe:1965}
\begin{equation}\label{Sutcliffe pressure frames}
p = \gamma^2\tilde{p} \qquad\textrm{(Sutcliffe)}
\end{equation}
Our aim is to establish a rigorous theoretical framework for the relativistic thermodynamics of a perfect fluid based on the Lorentz invariance between an inertial frame $\mathcal{R}$ and a local rest frame $\mathcal{\tilde{R}}$. Our main objective is to settle once for all the debate on the relativistic transformations laws of thermodynamic quantities by using a consistent and systematic approach. Another important purpose is to determine the structure of the relativistic continuity equations for extensive thermodynamic scalar and vector quantities. Such an approach is expected to lead to a clear identification of the relativistic corrections to the dynamics of a perfect fluid. Finally, our synthetic approach allows us to obtain elegant axiomatic statements of the relativistic first and second law of thermodynamics.\\ 

\noindent This publication is structured as follows : the structure of the relativistic continuity equations for scalar and vector quantities is established in Sec.~\ref{Relativistic continuity equations} by following the approach outlined by Stückelberg.~\cite{Stueckelberg:2013} The relativistic statements of the second and first law are presented in Sec.~\ref{Relativistic second law of thermodynamics} and~\ref{Relativistic first law of thermodynamics} together with the relativistic continuity equations for the entropy and energy-momentum. In Sec.~\ref{Relativistic orbital angular momentum continuity equation}, we establish the relativistic continuity equation for the orbital angular momentum, which implies the symmetry of the stress-energy-momentum tensor. The relativistic matter continuity equation is derived in Sec.~\ref{Relativistic matter continuity equation}. As a prelude to the study of the dynamics and thermodynamics of a perfect fluid, we first examine the dynamics of a relativistic dust in Sec.~\ref{Dynamics of a relativistic dust} and derive the corresponding stress-energy-momentum tensor in Sec.~\ref{Stress-energy-momentum tensor of a relativistic dust}. Taking into account the pressure and the internal energy density, we generalise our analysis to the study of the dynamics of a perfect fluid in Sec.~\ref{Dynamics of a relativistic perfect fluid} and derive the corresponding stress-energy-momentum tensor in Sec.~\ref{Stress-energy-momentum tensor of a relativistic perfect fluid}. This leads to the introduction of a relativistic inertia density that differs from the mass density. Using the thermodynamic definition of the temperature, the pressure and the chemical potential, we establish rigorously the relativistic transformation laws for these three intensive quantities in Sec.~\ref{Relativistic temperature},~\ref{Relativistic chemical potential} and~\ref{Relativistic pressure} respectively. The relativistic force density is established in Sec.~\ref{Relativistic force density}. The relativistic continuity equations for the energy and the momentum are derived from the relativistic continuity equation for the energy-momentum in Sec.~\ref{Relativistic energy continuity equation} and~\ref{Relativistic momentum continuity equation}, and the relativistic corrections are highlighted. In Sec.~\ref{Relativistic inertia continuity equation}, we derive a relativistic continuity equation for the inertia and show how and why it differs from the continuity equation for the mass that is usually assumed to hold. We derive the relativistic Euler equation from the relativistic continuity equations for the momentum and the inertia in Sec.~\ref{Relativistic Euler equation} and highlight the relativistic corrections. Finally, we conclude our analysis of the relativistic thermodynamics of a perfect fluid in Sec.~\ref{Conclusion}. In Appendices~\ref{Relativistic kinematics} and~\ref{Relativistic dynamics}, we present the fundamental concepts of relativistic kinematics and dynamics of a particle, which are the cornerstone of a relativistic theory.


\section{Relativistic continuity equations}
\label{Relativistic continuity equations}

\noindent We establish in this section the structure of the relativistic continuity equations for a frame-independent scalar quantity $F$ and for the contravariant component $F^{\nu}$ of a vector in Minkowski space. We follow the covariant approach based on the divergence theorem in Minkowski space outlined by Stückelberg.~\cite{Stueckelberg:2013} In order to do so, we consider a thermodynamic system within a fixed enclosure. An arbitrary extensive physical property is described by an extensive scalar function $F\left(t\right)$ of time $t$ in an inertial frame $\mathcal{R}$. By extensivity, the scalar function $F\left(t\right)$ is obtained by integrating the density $f\left(t,\boldsymbol{x}\right)$ over the spatial volume $V_3\left(t\right)$ of the system,~\cite{Brechet:2019}
\begin{equation}\label{function F 0}
F\left(t\right) = \int_{V_3\left(t\right)}dV_3\left(\boldsymbol{x}\right)f\left(t,\boldsymbol{x}\right)
\end{equation}
where $\boldsymbol{x}$ is the position spatial vector, $t$ is the time, $dV_3\left(\boldsymbol{x}\right)$ is the spatial infinitesimal volume in the inertial frame $\mathcal{R}$. To describe the time evolution of the extensive scalar function $F\left(t\right)$ in a relativistic manner, we define a covariant vector representing an infinitesimal spatial volume that is written in components in the inertial frame $\mathcal{R}$ as,
\begin{equation}\label{infinitesimal spatial covariant volume}
dV_{3\,\mu}\left(\boldsymbol{x}\right) = dV_3\left(\boldsymbol{x}\right)\delta_{0\mu}
\end{equation}
The covariant time component of the infinitesimal spatial volume vector~\eqref{infinitesimal spatial covariant volume} in the inertial frame $\mathcal{R}$ reads,
\begin{equation}\label{infinitesimal spatial covariant volume time}
dV_{3\,0}\left(\boldsymbol{x}\right) = dV_3\left(\boldsymbol{x}\right)
\end{equation}
We introduce a current density vector of the physical property $F\left(t\right)$ of contravariant components $j_f^{\mu}\left(t,\boldsymbol{x}\right)$ in the inertial frame $\mathcal{R}$. In the local rest frame $\mathcal{\tilde{R}}$, the contravariant component of the current density vector $\tilde{j}_f^{\mu}\left(t,\boldsymbol{x}\right)$ is defined as a purely temporal vector,
\begin{equation}\label{current density rest}
\tilde{j}_f^{\mu}\left(\tilde{t},\boldsymbol{\tilde{x}}\right) = c\,\tilde{f}\left(\tilde{t},\boldsymbol{\tilde{x}}\right)\delta^{\mu 0}
\end{equation}
where $\tilde{f}\left(\tilde{t},\boldsymbol{\tilde{x}}\right)$ is the density of the frame independent physical property $\tilde{F}\left(\tilde{t}\right)$ in the local rest frame $\mathcal{\tilde{R}}$,
\begin{equation}\label{F frame independent}
\boxed{F\left(t\right) = \tilde{F}\left(\tilde{t}\right)}
\end{equation}
Thus, the contravariant time component of the current density vector $\tilde{j}_f^{\mu}\left(t,\boldsymbol{x}\right)$ in the local rest frame $\mathcal{\tilde{R}}$ is written as,
\begin{equation}\label{temporal f contravariant}
\tilde{j}_f^0\left(\tilde{t},\boldsymbol{\tilde{x}}\right) = c\,\tilde{f}\left(\tilde{t},\boldsymbol{\tilde{x}}\right)
\end{equation}
In view of the contravariant component of the current density vector~\eqref{current density rest}, the contravariant components of the current density vector $j_f^{\mu}\left(\{x^{\rho}\}\right)$ in an inertial frame $\mathcal{R}$ are related to the contravariant components of the current density vector $\tilde{j}_f^{\nu}\left(\{\tilde{x}^{\rho}\}\right)$ in the local rest frame $\mathcal{\tilde{R}}$ by an inverse Lorentz transformation,
\begin{equation}\label{current density Lorentz}
j_f^{\mu}\left(\{x^{\rho}\}\right) = \left(\Lambda^{-1}\right)^{\mu}_{\phantom{\mu}\nu}\,\tilde{j}_f^{\nu}\left(\{\tilde{x}^{\rho}\}\right) = \left(\Lambda^{-1}\right)^{\mu}_{\phantom{\mu}0}\,c\,\tilde{f}\left(\{\tilde{x}^{\rho}\}\right)
\end{equation}
In view of the component of the Lorentz transformation~\eqref{Lorentz transformation inverse components} and the contravariant component of the current density vector~\eqref{current density Lorentz}, the contravariant time component of the current density vector is given by,
\begin{equation}\label{current density Lorentz time}
j_f^{0}\left(\{x^{\rho}\}\right) = \left(\Lambda^{-1}\right)^{0}_{\phantom{0}0}\,c\,\tilde{f}\left(\{\tilde{x}^{\rho}\}\right) = \gamma\,c\,\tilde{f}\left(\{\tilde{x}^{\rho}\}\right)
\end{equation}
and the contravariant spatial components of the current density vector are given by,
\begin{equation}\label{current density Lorentz space}
j_f^{j}\left(\{x^{\rho}\}\right) = \left(\Lambda^{-1}\right)^{j}_{\phantom{i}0}\,c\,\tilde{f}\left(\{\tilde{x}^{\rho}\}\right) = \gamma\,v^i\,\tilde{f}\left(\{\tilde{x}^{\rho}\}\right)\delta^{j}_{i}
\end{equation}
In view of the contravariant temporal and spatial components of the velocity vector~\eqref{temporal velocity contravariant} and~\eqref{spatial velocity contravariant and covariant}, and the contravariant temporal and spatial components of the current density vector~\eqref{current density Lorentz time} and~\eqref{current density Lorentz space}, the contravariant components of the current density in the inertial frame $\mathcal{R}$ are given by,
\begin{equation}\label{current density}
\boxed{j_f^{\mu}\left(\{x^{\rho}\}\right) = \tilde{f}\left(\{\tilde{x}^{\rho}\}\right)u^{\mu}\left(\{x^{\rho}\}\right)}
\end{equation}
In view of the contravariant time component of the current density vector~\eqref{temporal f contravariant} in the local rest frame $\mathcal{\tilde{R}}$, the contravariant time component of the current density vector in the inertial frame $\mathcal{R}$ is written as,
\begin{equation}\label{current density Lorentz time inertial}
j_f^{0}\left(\{x^{\rho}\}\right) = c\,f\left(\{x^{\rho}\}\right)
\end{equation}
Identifying the contravariant time components of the current density vector~\eqref{current density Lorentz time} and~\eqref{current density Lorentz time inertial}, we deduce that the density $f\left(t,\boldsymbol{x}\right)$ in the inertial frame $\mathcal{R}$ is related to the density $\tilde{f}\left(\tilde{t},\boldsymbol{\tilde{x}}\right)$ in the local rest frame $\mathcal{\tilde{R}}$ by,
\begin{equation}\label{density f frames}
\boxed{f\left(t,\boldsymbol{x}\right) = \gamma\,\tilde{f}\left(\tilde{t},\boldsymbol{\tilde{x}}\right)}
\end{equation}
According to the covariant time component of the infinitesimal spatial volume~\eqref{infinitesimal spatial covariant volume time} and the contravariant time component of the current density vector~\eqref{current density Lorentz time inertial}, the integrand of the integral relation~\eqref{function F 0} for the scalar quantity $F$ can be recast in the inertial frame $\mathcal{R}$ as,
\begin{equation}\label{integrand inertial frame}
dV_3\left(\boldsymbol{x}\right)f\left(t,\boldsymbol{x}\right) = \frac{1}{c}\,dV_{3\,0}\left(\boldsymbol{x}\right)\,j_f^{0}\left(t,\boldsymbol{x}\right)
\end{equation}
According to the covariant components~\eqref{infinitesimal spatial covariant volume} and the covariant time component~\eqref{infinitesimal spatial covariant volume time} of the infinitesimal spatial volume vector,
\begin{equation}\label{integrand inertial frame bis}
dV_{3\,\mu}\left(\{x^{\rho}\}\right)\,j_f^{\mu}\left(\{x^{\rho}\}\right) = dV_{3}\left(\boldsymbol{x}\right)\,\delta_{0\mu}\ j_f^{\mu}\left(t,\boldsymbol{x}\right) = dV_{3\,0}\left(\boldsymbol{x}\right)\,j_f^{0}\left(t,\boldsymbol{x}\right)
\end{equation}
In view of the contracted identity~\eqref{integrand inertial frame bis}, the integrand~\eqref{integrand inertial frame} reduces to,
\begin{equation}\label{integrand inertial frame ter}
dV_3\left(\boldsymbol{x}\right)f\left(t,\boldsymbol{x}\right) = \frac{1}{c}\,dV_{3\,\mu}\left(\{x^{\rho}\}\right)\,j_f^{\mu}\left(\{x^{\rho}\}\right)
\end{equation}
which is a frame independent scalar since it is the contraction of two vectors. Time $t$ in the inertial frame $\mathcal{R}$ defines a three dimensional spatial hypersurface $\Sigma_3\left(t\right)$ orthogonal to the worldline $\Sigma_1$. The volume of the system is a subset of this hypersurface, i.e. $V_3\left(t\right) \subset\Sigma_3\left(t\right)$. In view of the integrand~\eqref{integrand inertial frame ter}, the extensive scalar function~\eqref{function F 0} can be recast as,
\begin{equation}\label{function F}
F\left(t\right) = \frac{1}{c}\,\int_{V_3\left(t\right)}dV_{3\,\mu}\left(\{x^{\rho}\}\right)j^{\mu}_{f}\left(\{x^{\rho}\}\right)
\end{equation}
In order to determine the continuity equation of the scalar function $F\left(t\right)$, we write the variation of this function between during a time interval $[t_i,t_f]$,
\begin{align}\label{variation function F}
&\Delta F_{i\rightarrow f} \equiv F\left(t_f\right) -\,F\left(t_i\right) = \frac{1}{c}\,\int_{V_3\left(t_f\right)}dV_{3\,\mu}\left(\{x^{\rho}\}\right)j^{\mu}_{f}\left(\{x^{\rho}\}\right)\nonumber\\
&\phantom{\Delta F_{i\rightarrow f} \equiv F\left(t_f\right) -\,F\left(t_i\right) =} -\,\frac{1}{c}\,\int_{V_3\left(t_i\right)}dV_{3\,\mu}\left(\{x^{\rho}\}\right)j^{\mu}_{f}\left(\{x^{\rho}\}\right)
\end{align}
\begin{figure}[h]
\begin{center}
\includegraphics[width=0.75\textwidth]{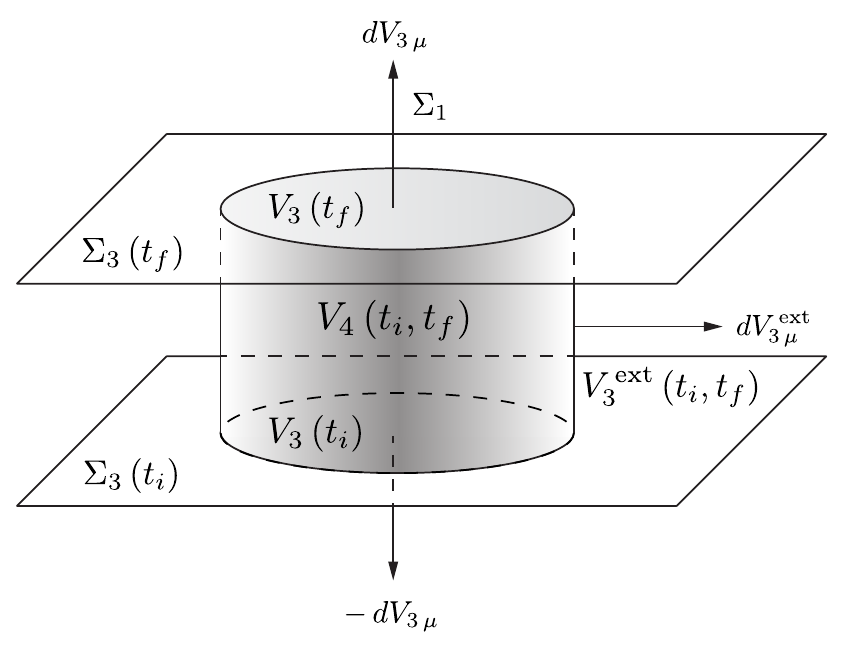}
\end{center}
\caption{Worldsheet of the system of hypervolume $V_4$ during the time interval $\left[t_i,t_f\right]$ in space-time represented by a cylinder. At the initial time $t_i$, the system is represented by a circular section of spatial volume $V_3\left(t_i\right)$. At the final time $t_f$, the system is represented by a circular section of spatial volume $V_3\left(t_f\right)$.}
\label{Worldsheet}
\end{figure}

\noindent The spatial volumes $V_3\left(t_i\right)$ and $V_3\left(t_f\right)$ of the system are represented in space-time by two sections of the parallel spatial hypersurfaces $\Sigma_3\left(t_i\right)$ and $\Sigma_3\left(t_f\right)$ (Fig.~\ref{Worldsheet}). The worldsheet of the fixed enclosure of the system defines a hypersurface of volume $V_3^{\,\text{ext}}\left(t_i,t_f\right)$ connecting the spatial volumes $V_3\left(t_i\right)$ and $V_3\left(t_f\right)$. We define an infinitesimal volume vector of components $dV^{\,\text{ext}}_{3\,\mu}\left(\{x^{\rho}\}\right)$ locally orthogonal to the hypersurface of volume $V_3^{\,\text{ext}}\left(t_i,t_f\right)$ containing the velocity vector $u^{\mu}$ associated to the worldline of the local rest frame. The orthogonality condition between the volume vector and the worldline is written in the inertial frame $\mathcal{R}$ as,
\begin{equation}\label{orthogonality condtion volume vectors}
dV^{\,\text{ext}}_{3\,\mu}\left(\{x^{\rho}\}\right)u^{\mu} = 0
\end{equation}
According to contravariant component of the current density vector~\eqref{current density} and the null identity~\eqref{orthogonality condtion volume vectors}, we obtain the orthogonality condition,
\begin{equation}\label{volume vector current density rest}
dV^{\,\text{ext}}_{3\,\mu}\left(\{x^{\rho}\}\right)j_f^{\mu}\left(\{x^{\rho}\}\right) = dV^{\,\text{ext}}_{3\,\mu}\left(\{x^{\rho}\}\right)\tilde{f}\left(\{\tilde{x}^{\rho}\}\right)u^{\mu}\left(\{x^{\rho}\}\right) = 0
\end{equation}
Integrating the orthogonality condition~\eqref{volume vector current density rest}, we obtain the null identity,
\begin{equation}\label{null integral acausal volume}
\frac{1}{c}\,\int_{V_3^{\,\text{ext}}\left(t_i,t_f\right)}dV^{\,\text{ext}}_{3\,\mu}\left(\{x^{\rho}\}\right)j^{\mu}_{f}\left(\{x^{\rho}\}\right) = 0
\end{equation}
In view of relation~\eqref{null integral acausal volume}, the variation of the scalar function $F$~\eqref{function F} during the time interval $[t_i,t_f]$ is recast as,
\begin{align}\label{variation function F bis}
&\Delta F_{i\rightarrow f} = \frac{1}{c}\,\int_{V_3\left(t_f\right)}dV_{3\,\mu}\left(\{x^{\rho}\}\right)j^{\mu}_{f}\left(\{x^{\rho}\}\right) + \frac{1}{c}\,\int_{V_3\left(t_i\right)}\left(-\,dV_{3\,\mu}\left(\{x^{\rho}\}\right)\right)j^{\mu}_{f}\left(\{x^{\rho}\}\right)\nonumber\\
&\phantom{\Delta F_{i\rightarrow f} =} + \frac{1}{c}\,\int_{V_3^{\,\text{ext}}\left(t_i,t_f\right)}dV^{\,\text{ext}}_{3\,\mu}\left(\{x^{\rho}\}\right)j^{\mu}_{f}\left(\{x^{\rho}\}\right)
\end{align}
The four-dimensional worldsheet of the system during the time interval $[t_i,t_f]$ has a hypervolume $V_4$.~\cite{Stueckelberg:2013} The boundary of the worldsheet of the system of hypervolume $V_4$ consists of the three-dimensional volumes $V_3\left(t_i\right)$, $V_3\left(t_f\right)$ and $V^{\,\text{ext}}\left(t_i,t_f\right)$. The infinitesimal volume vectors $dV_{3\,\mu}$, $-\,dV_{3\,\mu}$ and $dV^{\,\text{ext}}_{3\,\mu}$ are orthogonal to the boundaries and oriented outside the four-dimensional worldsheet of the system. Thus, the variation~\eqref{variation function F bis} of the scalar function $F$ during the time interval $[t_i,t_f]$ is recast as a closed integral over the boundary of the four-dimensional worldsheet of the system during this time interval,
\begin{equation}\label{variation function F ter}
\Delta F_{i\rightarrow f} = \frac{1}{c}\,\oint_{\partial V_4\left(t_i,t_f\right)}dV_{3\,\mu}\left(\{x^{\rho}\}\right)j^{\mu}_{f}\left(\{x^{\rho}\}\right)
\end{equation}
According to the divergence theorem,~\cite{Stueckelberg:2013} the integral on the three dimensional closed boundary of worldsheet of the system during the time interval $[t_i,t_f]$ is equal to the integral of the divergence of the integrand on the four dimensional worldsheet of the system during this time interval,
\begin{equation}\label{variation function F divergence}
\Delta F_{i\rightarrow f} = \frac{1}{c}\,\int_{V_4\left(t_i,t_f\right)}dV_4\left(\{x^{\rho}\}\right)\partial_{\mu}\,j^{\mu}_{f}\left(\{x^{\rho}\}\right)
\end{equation}
If the scalar function $F$ represents a conserved physical property, there is no variation~\eqref{variation function F divergence} during the time interval $[t_i,t_f]$, i.e. $\Delta F_{i\rightarrow f} = 0$. Otherwise, there is scalar function $\Sigma_{F\,i\rightarrow f}$ representing the physical source of $F$ in the system,
\begin{equation}\label{relativistic balance equation F}
\Delta F_{i\rightarrow f} = \Sigma_{F\,i\rightarrow f}
\end{equation}
which is the relativistic balance equation of $F$. The physical source of $F$ in the system is the integral of the scalar source density $\sigma_f$ of the scalar function density $f$ over the worldsheet of the system during the time interval $[t_i,t_f]$ divided by $c$,
\begin{equation}\label{variation function F physical cause int}
\Sigma_{F\,i\rightarrow f} = \frac{1}{c}\,\int_{V_4\left(t_i,t_f\right)}dV_4\left(\{x^{\rho}\}\right)\sigma_f\left(\{x^{\rho}\}\right)
\end{equation}
In view of the integral relations~\eqref{variation function F divergence} and~\eqref{variation function F physical cause int} for the variation of the scalar function $F$, the relativistic balance equation~\eqref{relativistic balance equation F} multiplied by $c$ yields,
\begin{equation}\label{relativistic balance equation F integral}
\int_{V_4\left(t_i,t_f\right)}dV_4\left(\{x^{\rho}\}\right)\partial_{\mu}\,j^{\mu}_{f}\left(\{x^{\rho}\}\right) = \int_{V_4\left(t_i,t_f\right)}dV_4\left(\{x^{\rho}\}\right)\sigma_f\left(\{x^{\rho}\}\right)
\end{equation}
The identification of the integrands in the relativistic balance equation~\eqref{relativistic balance equation F integral} yields the relativistic continuity equation of $F$,
\begin{equation}\label{relativistic continuity equation F}
\boxed{\partial_{\mu}\,j^{\mu}_{f}\left(\{x^{\rho}\}\right) = \sigma_f\left(\{x^{\rho}\}\right)}
\end{equation}
In view of the contravariant components of the current density~\eqref{current density}, the relativistic continuity equation is recast as,~\cite{Stueckelberg:2013}
\begin{equation}\label{relativistic continuity equation F bis}
\boxed{\partial_{\mu}\Big(\tilde{f}\left(\{\tilde{x}^{\rho}\}\right)u^{\mu}\left(\{x^{\rho}\}\right)\Big) = \sigma_f\left(\{x^{\rho}\}\right) \qquad \left(\textrm{inertial frame}\right)}
\end{equation}
The covariant time component of the partial derivative vectorial operator is written is written as,
\begin{equation}\label{temporal partial derivative contravariant}
\partial_0 = \frac{1}{c}\,\partial_t
\end{equation}
The gradient $\boldsymbol{\nabla}$ is expressed in covariant components in the orthonormal dual basis frame as,
\begin{equation}\label{gradient}
\boldsymbol{\nabla} = \boldsymbol{e}^{j}\,\partial_{j}
\end{equation}
Using the contravariant components of the velocity vector~\eqref{spatial velocity contravariant and covariant}, the contravariant components of the current density vector~\eqref{current density} and the relativistic transformation law for the density~\eqref{density f frames}, the contravariant spatial components of the current density vector are given by,
\begin{equation}\label{current density space}
j_f^{j}\left(\{x^{\rho}\}\right) = \gamma\,\tilde{f}\left(\tilde{t},\boldsymbol{\tilde{x}}\right)\,v^j\left(t,\boldsymbol{x}\right) = f\left(t,\boldsymbol{x}\right)\,v^j\left(t,\boldsymbol{x}\right)
\end{equation}
In view of the frame duality~\eqref{duality space}, the spatial velocity vector~\eqref{spatial velocity}, the contravariant temporal and spatial components of the current density vector~\eqref{current density Lorentz time inertial} and~\eqref{current density space}, the partial time derivative~\eqref{temporal partial derivative contravariant} and the gradient~\eqref{gradient}, the divergence of the current density vector is written as,
\begin{equation}\label{gradient current density}
\begin{split}
&\partial_{\mu}\,j^{\mu}_{f}\left(\{x^{\rho}\}\right) = \partial_{0}\,j^{0}_{f}\left(\{x^{\rho}\}\right) + \partial_{j}\,j^{j}_{f}\left(\{x^{\rho}\}\right)\\
&\phantom{\partial_{\mu}\,j^{\mu}_{f}\left(\{x^{\rho}\}\right)} = \partial_t\,f\left(t,\boldsymbol{x}\right) + \left(\boldsymbol{e}^{k}\,\partial_{k}\right)\cdot\Big(f\left(t,\boldsymbol{x}\right)v^j\left(t,\boldsymbol{x}\right)\boldsymbol{e}_{j}\Big)\\
&\phantom{\partial_{\mu}\,j^{\mu}_{f}\left(\{x^{\rho}\}\right)} = \partial_t\,f\left(t,\boldsymbol{x}\right) + \boldsymbol{\nabla}\cdot\Big(f\left(t,\boldsymbol{x}\right)\boldsymbol{v}\left(t,\boldsymbol{x}\right)\Big)
\end{split}
\end{equation}
In view of the divergence of the current density vector~\eqref{gradient current density}, the relativistic continuity equation~\eqref{relativistic continuity equation F} is written as,
\begin{equation}\label{relativistic continuity equation F split}
\boxed{\partial_t\,f\left(t,\boldsymbol{x}\right) + \boldsymbol{\nabla}\cdot\Big(f\left(t,\boldsymbol{x}\right)\boldsymbol{v}\left(t,\boldsymbol{x}\right)\Big) = \sigma_f\left(t,\boldsymbol{x}\right) \qquad \left(\textrm{inertial frame}\right)}
\end{equation}
and in view of the contravariant components~\eqref{current density space}, the current density spatial vector in the inertial frame $\mathcal{R}$ is given by,
\begin{equation}\label{spatial current density}
\boldsymbol{j}_f\left(t,\boldsymbol{x}\right) = f\left(t,\boldsymbol{x}\right)\boldsymbol{v}\left(t,\boldsymbol{x}\right)
\end{equation}
According to the relativistic continuity equation~\eqref{relativistic continuity equation F}, the source density of $f$ is a frame independent scalar,
\begin{equation}\label{source density f frame independent}
\boxed{\sigma_{f}\left(\{x^{\rho}\}\right) = \tilde{\sigma}_{\tilde{f}}\left(\{\tilde{x}^{\rho}\}\right)}
\end{equation}
In the local rest frame $\mathcal{\tilde{R}}$, in view of relations~\eqref{current density rest} and~\eqref{source density f frame independent}, the relativistic continuity equation~\eqref{relativistic continuity equation F bis} becomes,
\begin{equation}\label{relativistic continuity equation F ter}
\boxed{\tilde{\partial}_{\mu}\Big(\tilde{f}\left(\{\tilde{x}^{\rho}\}\right)\tilde{u}^{\mu}\left(\{\tilde{x}^{\rho}\}\right)\Big) = \tilde{\sigma}_{f}\left(\{\tilde{x}^{\rho}\}\right) \qquad \left(\textrm{local rest frame}\right)}
\end{equation}
Using the identities,
\begin{equation}\label{relativistic identities derivative velocity}
\tilde{u}^{\mu} = c\,\delta^{0\mu} \qquad\textrm{and}\qquad \tilde{\partial}_{\mu} = c^{-1}\,\partial_{\tilde{t}}
\end{equation}
the relativistic continuity equation~\eqref{relativistic continuity equation F ter} in the local rest frame $\mathcal{\tilde{R}}$ becomes,
\begin{equation}\label{relativistic continuity equation F quad}
\boxed{\partial_{\tilde{t}}\,\tilde{f}\left(\tilde{t},\boldsymbol{\tilde{x}}\right) = \tilde{\sigma}_{\tilde{f}}\left(\tilde{t},\boldsymbol{\tilde{x}}\right) \qquad \left(\textrm{local rest frame}\right)}
\end{equation}
The extension of the current approach to determine the relativistic continuity equation of a vector function is straightforward. By analogy with the scalar function~\eqref{function F} of time $F\left(t\right)$, the contravariant scalar components $F^{\nu}\left(t\right)$ of the vector function, where $\nu \in\{0,1,2,3\}$, are written as,~\cite{Stueckelberg:2013}
\begin{equation}\label{function F vec}
F^{\nu}\left(t\right) = \frac{1}{c}\,\int_{V_3\left(t\right)}dV_{3\,\mu}\left(\{x^{\rho}\}\right)j^{\mu\nu}_{f}\left(\{x^{\rho}\}\right)
\end{equation}
where $j^{\mu\nu}_{f}\left(\{x^{\rho}\}\right)$ are the components of the current density of the physical property $F^{\nu}$. The current density of a scalar (rank-0 tensor) property $F$ is a vector (rank-1 tensor) of components $j^{\mu}$ and the current density of a vector (rank-1 tensor) property of components $F^{\nu}$ is a rank-2 tensor of components $j^{\mu\nu}$. By analogy with the scalar function~\eqref{function F} of time $F\left(t\right)$, the variation of the scalar components $F^{\nu}\left(t\right)$ of a vector function is written as,
\begin{equation}\label{variation function F vec divergence}
\Delta F^{\nu}_{i\rightarrow f} = \frac{1}{c}\,\int_{V_4\left(t_i,t_f\right)}dV_4\left(\{x^{\rho}\}\right)\partial_{\mu}\,j^{\mu\nu}_{f}\left(\{x^{\rho}\}\right)
\end{equation}
If the vector function~\eqref{variation function F vec divergence} of scalar components $F^{\nu}$ represents a conserved physical property, there is no variation during the time interval $[t_i,t_f]$, i.e. $\Delta F^{\nu}_{i\rightarrow f} = 0$. Otherwise, there is scalar function $\Sigma^{\nu}_{F\,i\rightarrow f}$ representing the physical source of $F^{\nu}$ in the system,
\begin{equation}\label{relativistic balance equation F vec}
\Delta F^{\nu}_{i\rightarrow f} = \Sigma^{\nu}_{F\,i\rightarrow f}
\end{equation}
which is the component of relativistic balance equation of the vector function. The physical source of $F^{\nu}$ in the system is the integral of the component $\sigma^{\nu}_f$ of the vector source density of the component $f^{\nu}$ of the vector function density over the worldsheet of the system during the time interval $[t_i,t_f]$ divided by $c$,
\begin{equation}\label{variation function F vec physical cause int}
\Sigma^{\nu}_{F\,i\rightarrow f} = \frac{1}{c}\,\int_{V_4\left(t_i,t_f\right)}dV_4\left(\{x^{\rho}\}\right)\sigma^{\nu}_f\left(\{x^{\rho}\}\right)
\end{equation}
In view of the integral relations~\eqref{variation function F vec divergence} and~\eqref{variation function F vec physical cause int}, the relativistic balance equation~\eqref{relativistic balance equation F vec} multiplied by $c$ yields,
\begin{equation}\label{relativistic balance equation F vec integral}
\int_{V_4\left(t_i,t_f\right)}dV_4\left(\{x^{\rho}\}\right)\partial_{\mu}\,j^{\mu\nu}_{f}\left(\{x^{\rho}\}\right) = \int_{V_4\left(t_i,t_f\right)}dV_4\left(\{x^{\rho}\}\right)\sigma^{\nu}_f\left(\{x^{\rho}\}\right)
\end{equation}
The identification of the integrands in the relativistic balance equation~\eqref{relativistic balance equation F vec integral} yields the relativistic continuity equation of $F^{\nu}$,~\cite{Stueckelberg:2013}
\begin{equation}\label{relativistic continuity equation F vec}
\boxed{\partial_{\mu}\,j^{\mu\nu}_{f}\left(\{x^{\rho}\}\right) = \sigma^{\nu}_f\left(\{x^{\rho}\}\right)}
\end{equation}
%


\section{Relativistic second law of thermodynamics}
\label{Relativistic second law of thermodynamics}

\noindent We state in this section the relativistic first law of thermodynamics and establish the relativistic continuity equation of the frame-independent scalar function entropy $S$ based on the results of Sec.~\ref{Relativistic continuity equations}. In particular, we derive the transformation for the entropy density and show that the entropy source density is a frame independent quantity. On a microscopic scale, the elementary constituents of the fluid can be considered as hard spheres in local rest frame $\mathcal{\tilde{R}}$. Thus, the elementary constituents appear as ellipsoids flattened in the direction of motion in the inertial frame $\mathcal{R}$. The volume contraction in the direction of motion~\eqref{volume contraction} then ensures that the number $\Omega$ of microscopic states corresponding to a macroscopic state remains unchanged under a Lorentz transformation. Thus, in agreement with Einstein~\cite{Einstein:1907} and Planck~\cite{Planck:1908}, the entropy of the system is a frame independent quantity,
\begin{equation}\label{entropy S}
\boxed{S\left(t\right) = \tilde{S}\left(\tilde{t}\right)}
\end{equation}
The relativistic second law of thermodynamics reads :
\vspace{4mm}

\begin{tcolorbox}[colback=white]
For every thermodynamic system there exists a scalar extensive state function entropy $S\left(t\right)$. During the time interval $[t_i,t_f]$, where $t_i < t_f$, the entropy source $\Sigma_{S\,i\rightarrow f}$ is a monotonically increasing function of time $t$,
\begin{equation}\label{variation function S divergence}
\Delta S_{i\rightarrow f} = \Sigma_{S\,i\rightarrow f} \geqslant 0
\end{equation}
In the distant future, the entropy $S\left(t\right)$ of an isolated system tends to a finite maximum $S_{\,\text{max}}$ that is compatible with the constraints imposed on the system, such as its internal walls or characteristics of the enclosure,
\begin{equation}\label{S max}
\lim_{t \to \infty} S\left(t\right) = S_{\,\text{max}}
\end{equation}
\end{tcolorbox}

\noindent The contravariant components of the entropy current density vector~\eqref{current density} in the inertial frame $\mathcal{R}$ are written as,~\cite{Stueckelberg:2013}
\begin{equation}\label{temporal entropy contravariant}
\boxed{j_s^{\mu}\left(\{x^{\rho}\}\right) = \tilde{s}\left(\{\tilde{x}^{\rho}\}\right)u^{\mu}\left(\{x^{\rho}\}\right)}
\end{equation}
where $\tilde{s}\left(\{\tilde{x}^{\rho}\}\right)$ is the entropy density in the local rest frame $\mathcal{\tilde{R}}$. According to the relativistic transformation law for a density~\eqref{density f frames}, the entropy density $s\left(t,\boldsymbol{x}\right)$ in the inertial frame $\mathcal{R}$ is related to the entropy density $\tilde{s}\left(\tilde{t},\boldsymbol{\tilde{x}}\right)$ in the local rest frame $\mathcal{\tilde{R}}$ by,
\begin{equation}\label{density s frames}
\boxed{s\left(t,\boldsymbol{x}\right) = \gamma\,\tilde{s}\left(\tilde{t},\boldsymbol{\tilde{x}}\right)}
\end{equation}
The entropy~\eqref{function F} reads,
\begin{equation}\label{entropy}
S\left(t\right) = \frac{1}{c}\,\int_{V_3\left(t\right)}dV_{3\,\mu}\left(\{x^{\rho}\}\right)j^{\mu}_{s}\left(\{x^{\rho}\}\right)
\end{equation}
Using Gauss' theorem, the entropy variation~\eqref{variation function F divergence} during the time interval $[t_i,t_f]$ is written as,
\begin{equation}\label{entropy variation}
\Delta S_{i\rightarrow f} = \frac{1}{c}\,\int_{V_4\left(t_i,t_f\right)}dV_4\left(\{x^{\rho}\}\right)\partial_{\mu}\,j^{\mu}_{s}\left(\{x^{\rho}\}\right)
\end{equation}
The entropy source~\eqref{variation function F physical cause int} is written as,
\begin{equation}\label{entropy source}
\Sigma_{S\,i\rightarrow f} = \frac{1}{c}\,\int_{V_4\left(t_i,t_f\right)}dV_4\left(\{x^{\rho}\}\right)\sigma_s\left(\{x^{\rho}\}\right)
\end{equation}
In view of the second law of thermodynamics~\eqref{variation function S divergence}, the identification of the integrands of the entropy variation~\eqref{entropy variation} and the entropy source~\eqref{entropy source} yields the relativistic entropy continuity equation,
\begin{equation}\label{relativistic entropy continuity equation}
\boxed{\partial_{\mu}\,j^{\mu}_{s}\left(\{x^{\rho}\}\right) = \sigma_s\left(\{x^{\rho}\}\right) \geqslant 0}
\end{equation}
In view of the contravariant components of the entropy current density vector~\eqref{temporal entropy contravariant}, the relativistic entropy continuity equation~\eqref{relativistic entropy continuity equation} is recast as,~\cite{Stueckelberg:2013}
\begin{equation}\label{relativistic entropy continuity equation product}
\boxed{\partial_{\mu}\Big(\tilde{s}\left(\{\tilde{x}^{\rho}\}\right)u^{\mu}\left(\{x^{\rho}\}\right)\Big) = \sigma_s\left(\{x^{\rho}\}\right) \geqslant 0 \qquad \left(\textrm{inertial frame}\right)}
\end{equation}
The relativistic continuity equation for the entropy~\eqref{relativistic continuity equation F split} in the inertial frame $\mathcal{R}$ is written as,
\begin{equation}\label{relativistic entropy continuity equation bis}
\boxed{\partial_t\,s\left(t,\boldsymbol{x}\right) + \boldsymbol{\nabla}\cdot\Big(s\left(t,\boldsymbol{x}\right)\boldsymbol{v}\left(t,\boldsymbol{x}\right)\Big) = \sigma_s\left(t,\boldsymbol{x}\right) \geqslant 0 \qquad \left(\textrm{inertial frame}\right)}
\end{equation}
and the entropy current density spatial vector~\eqref{spatial current density} in the inertial frame $\mathcal{R}$ is given by,
\begin{equation}\label{entropy current density}
\boldsymbol{j}_s\left(t,\boldsymbol{x}\right) = s\left(t,\boldsymbol{x}\right)\boldsymbol{v}\left(t,\boldsymbol{x}\right)
\end{equation}
According to the relativistic continuity equation~\eqref{relativistic entropy continuity equation}, the entropy source density is a frame independent scalar,
\begin{equation}\label{entropy source density frame independent}
\boxed{\sigma_{s}\left(\{x^{\rho}\}\right) = \tilde{\sigma}_{\tilde{s}}\left(\{\tilde{x}^{\rho}\}\right)}
\end{equation}
In the local rest frame $\mathcal{\tilde{R}}$, in view of the entropy source density~\eqref{entropy source density frame independent} and the contravariant components of the entropy current density vector,
\begin{equation}\label{entropy current density rest}
\tilde{j}^{\mu}_{s}\left(\{\tilde{x}^{\rho}\}\right) = \tilde{s}\left(\{\tilde{x}^{\rho}\}\right)\tilde{u}^{\mu}\left(\{\tilde{x}^{\rho}\}\right)
\end{equation}
the relativistic continuity equation~\eqref{relativistic entropy continuity equation} becomes,
\begin{equation}\label{relativistic entropy continuity equation ter}
\boxed{\tilde{\partial}_{\mu}\Big(\tilde{s}\left(\{\tilde{x}^{\rho}\}\right)\tilde{u}^{\mu}\left(\{\tilde{x}^{\rho}\}\right)\Big) = \tilde{\sigma}_{\tilde{s}}\left(\{\tilde{x}^{\rho}\}\right) \geqslant 0 \qquad \left(\textrm{local rest frame}\right)}
\end{equation}
Using the identities~\eqref{relativistic identities derivative velocity}, the relativistic entropy continuity equation~\eqref{relativistic entropy continuity equation ter} in the local rest frame $\mathcal{\tilde{R}}$ becomes,
\begin{equation}\label{relativistic entropy continuity equation quad}
\boxed{\partial_{\tilde{t}}\,\tilde{s}\left(\tilde{t},\boldsymbol{\tilde{x}}\right) = \tilde{\sigma}_{\tilde{s}}\left(\tilde{t},\boldsymbol{\tilde{x}}\right) \geqslant 0 \qquad \left(\textrm{local rest frame}\right)}
\end{equation}
%


\section{Relativistic first law of thermodynamics}
\label{Relativistic first law of thermodynamics}

\noindent We state in this section the relativistic first law of thermodynamics and establish the relativistic continuity equation of the contravariant component $P^{\nu}$ of the momentum vector in Minkowski space based on the results of Sec.~\ref{Relativistic continuity equations}. In order to do so, we use the fact that the momentum source is proportional to the contravariant component $F^{\nu}$ of the net force exerted on the system. The relativistic first law of thermodynamics reads :
\vspace{4mm}

\begin{tcolorbox}[colback=white]

For every thermodynamic system there exists a vector extensive state function momentum of contravariant components $P^{\nu}$. During the time interval $[t_i,t_f]$, where $t_i < t_f$, the momentum source $\Sigma^{\nu}_{P\,i\rightarrow f}$ is the net force exerted on the system multiplied by the time interval $F^{\nu}\,\Delta t_{i\rightarrow f}$, 
\begin{equation}\label{variation function P divergence}
\Delta P^{\nu}_{i\rightarrow f} = F^{\nu}\,\Delta t_{i\rightarrow f}
\end{equation}

\end{tcolorbox}

\noindent For an isolated system, there is not net force acting on the system, i.e. $F^{\nu}_{i\rightarrow f}$. Thus the momentum variation during the time interval $[t_i,t_f]$ vanishes, i.e. $\Delta P^{\nu}_{i\rightarrow f} = 0$, which implies that the components of the momentum vector are conserved, 
\begin{equation}\label{variation function P divergence isolated}
\boxed{P^{\nu}\left(t\right) = \text{const} \qquad\text{(isolated system)}}
\end{equation}
The scalar components of the momentum vector~\eqref{function F vec} read,
\begin{equation}\label{momentum component}
P^{\nu}\left(t\right) = \frac{1}{c}\,\int_{V_3\left(t\right)}dV_{3\,\mu}\left(\{x^{\rho}\}\right)T^{\mu\nu}\left(\{x^{\rho}\}\right)
\end{equation}
where $T^{\mu\nu}$ are the contravariant components of the rank-2 stress-energy-momentum tensor. The momentum variation~\eqref{variation function F vec divergence} is written as,
\begin{equation}\label{momentum variation}
\Delta P^{\nu}_{i\rightarrow f} = \frac{1}{c}\,\int_{V_4\left(t_i,t_f\right)}dV_4\left(\{x^{\rho}\}\right)\partial_{\mu}\,T^{\mu\nu}\left(\{x^{\rho}\}\right)
\end{equation}
The net force~\eqref{variation function F vec physical cause int} exerted on the system multiplied by the time interval $F^{\nu}\,\Delta t_{i\rightarrow f}$ is written as,
\begin{equation}\label{momentum source}
F^{\nu}\,\Delta t_{i\rightarrow f} = \frac{1}{c}\,\int_{V_4\left(t_i,t_f\right)}dV_4\left(\{x^{\rho}\}\right)f^{\nu}\left(\{x^{\rho}\}\right)
\end{equation}
where $f^{\nu}$ is the component of the force density vector. In view of the first law of thermodynamics~\eqref{variation function P divergence}, the identification of the integrands in the momentum variation~\eqref{momentum variation} and the momentum source~\eqref{momentum source} yields the relativistic energy-momentum continuity equation,
\begin{equation}\label{relativistic momentum continuity equation}
\boxed{\partial_{\mu}\,T^{\mu\nu}\left(\{x^{\rho}\}\right) = f^{\nu}\left(\{x^{\rho}\}\right)}
\end{equation}
%


\section{Relativistic orbital angular momentum continuity equation}
\label{Relativistic orbital angular momentum continuity equation}

\noindent We show in this section that in the absence of intrinsic angular momentum, the relativistic continuity equation for the contravariant components of the orbital angular momentum tensor $L^{\mu\nu}$ and the relativistic energy-momentum continuity equation~\eqref{relativistic momentum continuity equation}, established in \S~\ref{Relativistic first law of thermodynamics}, imply that the contravariant components $T^{\mu\nu}$ of the stress energy momentum tensor are symmetric. The contravariant components of the orbital angular momentum tensor are defined as the antisymmetric product of the contravariant components of the position vector of the centre of mass and the momentum vector,
\begin{equation}\label{angular momentum 0}
L^{\mu\nu} = X^{\mu}\,P^{\nu} -\,X^{\nu}\,P^{\mu}
\end{equation}
In view of the contravariant components of the momentum vector~\eqref{momentum component} the contravariant components of the orbital angular momentum tensor~\eqref{angular momentum 0} are written as,~\cite{Schutz:2009,Weinberg:1972}
\begin{equation}\label{angular momentum}
L^{\mu\nu} = \frac{1}{c}\,\int_{V_3\left(t\right)}dV_{3\,\rho}\left(x^{\mu}\,T^{\rho\nu}-\,x^{\nu}\,T^{\rho\mu}\right)
\end{equation}
Using Gauss' theorem, the variation of the contravariant components of the orbital angular momentum tensor during the time interval $[t_i,t_f]$ are written as,
\begin{equation}\label{variation angular momentum}
\Delta L^{\mu\nu}_{i\rightarrow f} = \frac{1}{c}\,\int_{V_4\left(t_i,t_f\right)}dV_4\,\partial_{\rho}\left(x^{\mu}\,T^{\rho\nu}-\,x^{\nu}\,T^{\rho\mu}\right)
\end{equation}
The orbital angular momentum source is the torque,
\begin{equation}\label{angular momentum source}
\Sigma_{L^{\mu\nu}\,i\rightarrow f} = \frac{1}{c}\,\int_{V_4\left(t_i,t_f\right)}dV_4\,t^{\mu\nu}
\end{equation}
where the contravariant components of the torque density tensor are defined as,
\begin{equation}\label{relativistic torque density tensor}
t^{\mu\nu} = x^{\mu}\,f^{\nu} -\,x^{\nu}\,f^{\mu} 
\end{equation}
The relativistic orbital angular momentum theorem is written as,
\begin{equation}\label{angular momentum theorem}
\Delta L^{\mu\nu}_{i\rightarrow f} = \Sigma_{L^{\mu\nu}\,i\rightarrow f}
\end{equation}
In view of the angular momentum theorem~\eqref{angular momentum theorem}, the identification of the integrands in the angular momentum variation~\eqref{variation angular momentum} and the angular momentum source~\eqref{angular momentum source} yields the relativistic orbital angular momentum continuity equation,
\begin{equation}\label{relativistic angular momentum continuity equation}
\boxed{\partial_{\rho}\left(x^{\mu}\,T^{\rho\nu}-\,x^{\nu}\,T^{\rho\mu}\right) = t^{\mu\nu}}
\end{equation}
According to relativistic energy-momentum continuity equation~\eqref{relativistic momentum continuity equation} and the identity $\partial_{\rho}\,x^{\mu} = \delta_{\rho}^{\mu}$, the relativistic angular momentum continuity equation~\eqref{relativistic angular momentum continuity equation} yields the identity,  
\begin{equation}\label{relativistic angular momentum continuity equation bis}
T^{\mu\nu} -\,T^{\nu\mu} + x^{\mu}\,f^{\nu} -\,x^{\nu}\,f^{\mu} = t^{\mu\nu}
\end{equation}
In view of the relativistic torque density tensor~\eqref{relativistic torque density tensor}, the identity~\eqref{relativistic angular momentum continuity equation bis} implies that the stress-energy-momentum tensor is symmetric,~\cite{Weinberg:1972}
\begin{equation}\label{stress energy momentum tensor symmetric}
\boxed{T^{\mu\nu} = T^{\nu\mu}}
\end{equation}
provided there is no intrinsic angular momentum, i.e. $S^{\mu\nu} = 0$.
%


\section{Relativistic matter continuity equation}
\label{Relativistic matter continuity equation}

\noindent We establish in this section the relativistic continuity equation of the frame-independent scalar function number of moles of substance $N$ based on the results of \S~\ref{Relativistic continuity equations}. We also derive the relativistic transformation law for the number density. In order to do so, we consider a relativistic fluid consisting of a single substance. In special relativity, the number of moles of substance in a system is a frame independent quantity,
\begin{equation}\label{number N}
\boxed{N = \tilde{N}}
\end{equation}
The contravariant components of the current density vector of substance~\eqref{current density} is written as,~\cite{Stueckelberg:2013}
\begin{equation}\label{temporal matter contravariant}
j_{n}^{\mu} = \tilde{n}\,u^{\mu}
\end{equation}
where $\tilde{n}$ is the number density in the local rest frame $\mathcal{\tilde{R}}$. According to the relativistic transformation law for the number density~\eqref{density f frames}, the number density $n$ in the inertial frame $\mathcal{R}$ is related to the number density $\tilde{n}$ in the local rest frame $\mathcal{\tilde{R}}$ by,
\begin{equation}\label{density n frames}
\boxed{n = \gamma\,\tilde{n}}
\end{equation}
The number of moles of substance~\eqref{function F} reads,
\begin{equation}\label{number of moles}
N = \frac{1}{c}\,\int_{V_3\left(t\right)}dV_{3\,\mu}\,j^{\mu}_{n}
\end{equation}
Using Gauss' theorem, the variation of the number of moles of substance~\eqref{variation function F divergence} during the time interval $[t_i,t_f]$ is written as,
\begin{equation}\label{number of moles variation}
\Delta N_{i\rightarrow f} = \frac{1}{c}\,\int_{V_4\left(t_i,t_f\right)}dV_4\,\partial_{\mu}\,j^{\mu}_{n}
\end{equation}
There is no matter source~\eqref{variation function F physical cause int} for a relativistic perfect fluid consisting of a single substance,
\begin{equation}\label{matter source}
\Sigma_{N\,i\rightarrow f} = \frac{1}{c}\,\int_{V_4\left(t_i,t_f\right)}dV_4\,\sigma_{n} = 0
\end{equation}
The matter conservation equation for a relativistic perfect fluid is written as,
\begin{equation}\label{matter conservation}
\Delta N_{i\rightarrow f} = \Sigma_{N\,i\rightarrow f} = 0
\end{equation}
In view of the matter conservation equation~\eqref{matter conservation}, the identification of the integrands in the variation of the number of moles~\eqref{number of moles variation} and the matter source~\eqref{matter source} yields the relativistic matter continuity equation,~\cite{Stueckelberg:2013}
\begin{equation}\label{relativistic matter continuity equation}
\boxed{\partial_{\mu}\,j^{\mu}_{n} = 0}
\end{equation}
In view of the contravariant components of the current density vector of substance~\eqref{temporal matter contravariant}, the relativistic matter continuity equation~\eqref{relativistic matter continuity equation} is recast as,
\begin{equation}\label{relativistic matter continuity equation product}
\boxed{\partial_{\mu}\left(\tilde{n}\,u^{\mu}\right) = 0 \qquad \left(\textrm{inertial frame}\right)}
\end{equation}
The relativistic matter continuity equation~\eqref{relativistic continuity equation F split} in the inertial frame $\mathcal{R}$ is written as,
\begin{equation}\label{relativistic matter continuity equation bis}
\boxed{\partial_t\,n + \boldsymbol{\nabla}\cdot\left(n\,\boldsymbol{v}\right) = 0 \qquad \left(\textrm{inertial frame}\right)}
\end{equation}
and the spatial vector matter current density~\eqref{spatial current density} in the inertial frame $\mathcal{R}$ is given by,
\begin{equation}\label{matter current density}
\boldsymbol{j}_{n} = n\,\boldsymbol{v}
\end{equation}
In the local rest frame $\mathcal{\tilde{R}}$, in view of the contravariant components of the current density vector of substance,
\begin{equation}\label{matter current density rest}
\tilde{j}^{\mu}_{n} = \tilde{n}\,\tilde{u}^{\mu}
\end{equation}
the relativistic continuity equation~\eqref{relativistic matter continuity equation} becomes,
\begin{equation}\label{relativistic matter continuity equation ter}
\boxed{\tilde{\partial}_{\mu}\left(\tilde{n}\,\tilde{u}^{\mu}\right) = 0 \qquad \left(\textrm{local rest frame}\right)}
\end{equation}
Using the identities~\eqref{relativistic identities derivative velocity}, the relativistic matter continuity equation~\eqref{relativistic matter continuity equation ter} in the local rest frame $\mathcal{\tilde{R}}$ becomes,
\begin{equation}\label{relativistic matter continuity equation quad}
\boxed{\partial_{\tilde{t}}\,\tilde{n} = 0 \qquad \left(\textrm{local rest frame}\right)}
\end{equation}
%


\section{Dynamics of a relativistic dust}
\label{Dynamics of a relativistic dust}

\noindent In order to generalise the relativistic dynamics of a particle described in \S~\ref{Relativistic dynamics}, we consider in this section a relativistic dust, which is a continuous medium consisting of relativistic non interacting particles. In particular, we derive the relativistic transformation laws for the mass density and the energy density. In the local rest frame $\mathcal{\tilde{R}}$, the mass density of a relativistic dust is defined as,~\cite{Schutz:2009}
\begin{equation}\label{mass density rest frame}
\tilde{\rho} = \tilde{n}\,\tilde{M}
\end{equation}
In view of the mass density~\eqref{mass density rest frame} in the local rest frame $\mathcal{\tilde{R}}$, the mass density of a relativistic dust is defined in the inertial frame $\mathcal{R}$ as,
\begin{equation}\label{mass density inertial frame}
\rho = n\,M
\end{equation}
According to the relativistic transformation laws for the mass~\eqref{mass M frames} and the number density~\eqref{density n frames}, the mass density $\rho$ in the inertial frame $\mathcal{R}$ is expressed in terms of the mass density $\tilde{\rho}$ in the rest frame $\mathcal{\tilde{R}}$ as,~\cite{Hobson:2006}
\begin{equation}\label{density rho frames}
\boxed{\rho = \gamma^2\,\tilde{\rho}}
\end{equation}
which shows that under a Lorentz transformation the mass density transforms as a component of a rank-2 tensor and not as a component of a vector. In view of the momentum vector~\eqref{momentum contravariant}, the relativistic transformation law of the number density~\eqref{density n frames} and the mass density~\eqref{mass density rest frame}, the contravariant components of the momentum density vector of a relativistic dust in the inertial frame $\mathcal{R}$ are given by,
\begin{equation}\label{momentum density contravariant}
p^{\mu} = n\,P^{\mu} = n\,\tilde{M}\,u^{\mu} = \gamma\,\tilde{n}\,\tilde{M}\,u^{\mu} = \gamma\,\tilde{\rho}\,u^{\mu}
\end{equation}
In view of the covariant components of the velocity vector~\eqref{spatial velocity contravariant and covariant}, and the momentum vector~\eqref{momentum covariant}, the relativistic transformation law of the number density~\eqref{density n frames} and the mass density~\eqref{mass density rest frame}, the covariant components of the momentum density vector of a relativistic dust in the inertial frame $\mathcal{R}$ are written as,
\begin{equation}\label{momentum density covariant}
p_{\mu} = n\,P_{\mu} = n\,\tilde{M}\,u_{\mu} = \gamma\,\tilde{n}\,\tilde{M}\,u_{\mu} = \gamma\,\tilde{\rho}\,u_{\mu}
\end{equation}
In contrast to the momentum, the momentum density is not a vector since the contravariant~\eqref{momentum density contravariant} or covariant~\eqref{momentum density covariant} components of the momentum density vector are the product of the frame dependent factor $\gamma\,\tilde{\rho}$ and the contravariant $u^{\mu}$ or covariant $u_{\mu}$ components of the velocity vector. Using the energy~\eqref{energy rest} and the mass~\eqref{mass density rest frame}, the energy density $\tilde{e}$ in the local rest frame $\mathcal{\tilde{R}}$ is written as,
\begin{equation}\label{energy density rest}
\tilde{e} = \tilde{n}\,\tilde{E} = \tilde{n}\,\tilde{M}\,c^2 = \tilde{\rho}\,c^2
\end{equation}
Using the relativistic transformation laws for the energy~\eqref{energy E frames}, the number density~\eqref{density n frames} and the mass density~\eqref{density rho frames}, and the rest energy density~\eqref{energy density rest}, the energy density $\tilde{e}$ in the inertial frame $\mathcal{R}$ is written as,
\begin{equation}\label{energy density}
e = n\,E = \gamma^2\,\tilde{n}\,\tilde{E} = \gamma^2\,\tilde{n}\,\tilde{M}\,c^2 = \gamma^2\,\tilde{\rho}\,c^2 = \rho\,c^2
\end{equation}
The energy density~\eqref{energy density} in the inertial frame $\mathcal{R}$ is related to the energy density~\eqref{energy density rest} in the local rest frame $\mathcal{\tilde{R}}$ by,
\begin{equation}\label{density e frames}
\boxed{e = \gamma^2\,\tilde{e}}
\end{equation}
In view of the relativistic transformation law for the mass density~\eqref{density rho frames}, the contravariant components of the momentum density vector~\eqref{momentum density contravariant}, and the energy density~\eqref{energy density}, the contravariant time component of the momentum density in the inertial frame $\mathcal{R}$ is given by,
\begin{equation}\label{temporal momentum density contravariant}
p^{0} = \frac{e}{c} = \rho\,c = \gamma^2\tilde{\rho}\,c
\end{equation}
In view of the relativistic transformation law for the mass density~\eqref{density rho frames}, the covariant component of the momentum density vector~\eqref{momentum density covariant}, and the energy density~\eqref{energy density}, the covariant time component of the momentum density in the inertial frame $\mathcal{R}$ is written as,
\begin{equation}\label{temporal momentum density covariant}
p_{0} = -\,\frac{e}{c} = -\,\rho\,c = -\,\gamma^2\tilde{\rho}\,c
\end{equation}
According to the spatial velocity vector~\eqref{spatial velocity}, the relativistic transformation law for the mass density~\eqref{density rho frames} and the contravariant components of the momentum density vector~\eqref{momentum density contravariant}, the spatial momentum density vector in the inertial frame $\mathcal{R}$ is written as,
\begin{equation}\label{momentum density}
\boldsymbol{p} = p^j\,\boldsymbol{e}_{j} = \gamma^2\,\tilde{\rho}\,v^j\,\boldsymbol{e}_{j}  = \gamma^2\tilde{\rho}\,\boldsymbol{v} = \rho\,\boldsymbol{v}
\end{equation}
where $p^j\,\boldsymbol{e}_{j} = p_j\,\boldsymbol{e}^{j}$.


\section{Stress-energy-momentum tensor of a relativistic dust}
\label{Stress-energy-momentum tensor of a relativistic dust}

\noindent We determine the contravariant components $T^{\mu\nu}$ of a stress energy-momentum tensor for a relativistic dust. Since a relativistic dust is made of non interacting particles, the only non zero contravariant component of the stress-energy-momentum tensor of a relativistic dust in the local rest frame $\mathcal{\tilde{R}}$ is the contravariant time component defined as the rest energy density,~\cite{Schutz:2009}
\begin{equation}\label{stress-energy-momentum tensor 00 LRF}
\tilde{T}^{00} = \tilde{e} = \tilde{\rho}\,c^2
\end{equation}
Thus, the contravariant components of the stress-energy-momentum tensor of a relativistic dust are formally written as,
\begin{equation}\label{stress-energy-momentum tensor mu nu LRF 0}
\tilde{T}^{\mu\nu} = \tilde{\rho}\,c^2\,\delta^{\mu 0}\,\delta^{\nu 0}
\end{equation}
The contravariant components of the velocity vector are purely time-like in the local rest frame $\mathcal{R}$,
\begin{equation}\label{velocity LRF}
\tilde{u}^{\mu} = c\,\delta^{\mu 0}
\end{equation}
In view of the contravariant components of the velocity vector~\eqref{velocity LRF}, the contravariant components of the stress-energy-momentum tensor~\eqref{stress-energy-momentum tensor mu nu LRF 0} become,
\begin{equation}\label{stress-energy-momentum tensor mu nu LRF}
\boxed{\tilde{T}^{\mu\nu} = \tilde{\rho}\,\tilde{u}^{\mu}\,\tilde{u}^{\nu}}
\end{equation}
In view of the rest energy density~\eqref{stress-energy-momentum tensor 00 LRF}, the contravariant components of the stress-energy-momentum tensor~\eqref{stress-energy-momentum tensor mu nu LRF} are recast as,
\begin{equation}\label{stress-energy-momentum tensor mu nu fluid LRF second}
\boxed{\tilde{T}^{\mu\nu} = \frac{\tilde{e}}{c^2}\,\tilde{u}^{\mu}\,\tilde{u}^{\nu}}
\end{equation}
The contravariant components of the stress-energy-momentum tensor $T^{\mu\nu}$ of a relativistic dust in the inertial frame $\mathcal{R}$ are obtained by performing inverse Lorentz transformations on the contravariant components of the stress-energy-momentum tensor $\tilde{T}^{\mu\nu}$ of a relativistic dust in in the local rest frame $\mathcal{\tilde{R}}$,
\begin{equation}\label{stress-energy-momentum tensor mu nu Lorentz}
T^{\mu\nu} = \left(\Lambda^{-1}\right)^\mu_{\phantom{\mu}\rho}\left(\Lambda^{-1}\right)^\nu_{\phantom{\nu}\lambda}\,\tilde{T}^{\rho\lambda}
\end{equation}
Using the components of the stress-energy-momentum tensor~\eqref{stress-energy-momentum tensor mu nu LRF 0} in the local rest frame $\mathcal{\tilde{R}}$, the components of the stress-energy-momentum tensor~\eqref{stress-energy-momentum tensor mu nu Lorentz} in the inertial frame $\mathcal{R}$ are recast as,
\begin{equation}\label{stress-energy-momentum tensor mu nu Lorentz bis}
T^{\mu\nu} = \left(\Lambda^{-1}\right)^\mu_{\phantom{\mu}0}\left(\Lambda^{-1}\right)^\nu_{\phantom{0}0}\,\tilde{\rho}\,c^2
\end{equation}
In view of the components of the inverse Lorentz transformation~\eqref{Lorentz transformation inverse components} and the energy density~\eqref{energy density}, the contravariant time component of the stress-energy-momentum tensor~\eqref{stress-energy-momentum tensor mu nu Lorentz bis} of a relativistic dust is the energy density,
\begin{equation}\label{stress-energy-momentum 00}
T^{00} = \left(\Lambda^{-1}\right)^0_{\phantom{0}0}\left(\Lambda^{-1}\right)^0_{\phantom{0}0}\,\tilde{\rho}\,c^2 = \gamma^2\tilde{\rho}\,c^2 = e
\end{equation}
According to the contravariant time component of the velocity vector~\eqref{temporal velocity contravariant}, the contravariant time component of the stress-energy-momentum tensor~\eqref{stress-energy-momentum 00} is recast as,~\cite{Hobson:2006,Schutz:2009}
\begin{equation}\label{stress-energy-momentum 00 bis}
T^{00} = \tilde{\rho}\left(\gamma\,c\right)\left(\gamma\,c\right) = \tilde{\rho}\,u^{0}\,u^{0}
\end{equation}
In view of the components of the inverse Lorentz transformation~\eqref{Lorentz transformation inverse components} and the energy density~\eqref{energy density}, the contravariant temporal and spatial components of the stress-energy-momentum tensor~\eqref{stress-energy-momentum tensor mu nu Lorentz bis} of a relativistic dust are the components of the energy current density vector up to a factor $c^{-1}$,~\cite{Hobson:2006,Schutz:2009}
\begin{equation}\label{stress-energy-momentum j0}
T^{j0} = \left(\Lambda^{-1}\right)^j_{\phantom{j}0}\left(\Lambda^{-1}\right)^0_{\phantom{0}0}\,\tilde{\rho}\,c^2 = \gamma^2\tilde{\rho}\,c\,v^{j} = \frac{e}{c}\,v^j
\end{equation}
According to the contravariant temporal and spatial components of the velocity vector~\eqref{temporal velocity contravariant} and~\eqref{spatial velocity contravariant and covariant}, the contravariant temporal and spatial components of the stress-energy-momentum tensor~\eqref{stress-energy-momentum j0} are recast as,
\begin{equation}\label{stress-energy-momentum j0 bis}
T^{j0} = \tilde{\rho}\left(\gamma\,c\right)\left(\gamma\,v^{j}\right) = \tilde{\rho}\,u^{0}\,u^{j}
\end{equation}
In view of the components of the inverse Lorentz transformation~\eqref{Lorentz transformation inverse components} and the contravariant components of the spatial momentum density vector~\eqref{momentum density}, the contravariant space and time components of the stress-energy-momentum tensor~\eqref{stress-energy-momentum tensor mu nu Lorentz bis} of a relativistic dust are the components of the momentum density up to a factor $c$,~\cite{Hobson:2006,Schutz:2009}
\begin{equation}\label{stress-energy-momentum 0j}
T^{0j} = \left(\Lambda^{-1}\right)^0_{\phantom{0}0}\left(\Lambda^{-1}\right)^j_{\phantom{j}0}\,\tilde{\rho}\,c^2 = \gamma^2\tilde{\rho}\,c\,v^{j} = c\,p^j
\end{equation}
According to the contravariant temporal and spatial components of the velocity vector~\eqref{temporal velocity contravariant} and~\eqref{spatial velocity contravariant and covariant} and the contravariant of the momentum density vector~\eqref{momentum density contravariant}, the contravariant temporal and spatial components of the stress-energy-momentum tensor~\eqref{stress-energy-momentum 0j} are recast as,
\begin{equation}\label{stress-energy-momentum 0j bis}
T^{0j} = \tilde{\rho}\left(\gamma\,v^{j}\right)\left(\gamma\,c\right) = \tilde{\rho}\,u^{j}\,u^{0}
\end{equation}
In view of relation~\eqref{Lorentz transformation inverse components} and~\eqref{momentum density}, the contravariant space and time components of the stress-energy-momentum tensor~\eqref{stress-energy-momentum tensor mu nu Lorentz bis} of a relativistic dust are the components are the contravariant components of the momentum current density up to a factor $\gamma$,~\cite{Hobson:2006,Schutz:2009}
\begin{equation}\label{stress-energy-momentum jk}
T^{jk} = \left(\Lambda^{-1}\right)^j_{\phantom{j}0}\left(\Lambda^{-1}\right)^k_{\phantom{k}0}\,\tilde{\rho}\,c^2 = \gamma^2\tilde{\rho}\,v^{j}\,v^{k}
\end{equation}
According to the contravariant spatial components of the velocity vector~\eqref{spatial velocity contravariant and covariant}, the contravariant spatial components of the stress-energy-momentum tensor~\eqref{stress-energy-momentum jk} are recast as,
\begin{equation}\label{stress-energy-momentum jk bis}
T^{jk} = \tilde{\rho}\left(\gamma\,v^{j}\right)\left(\gamma\,v^{k}\right) = \tilde{\rho}\,u^{j}\,u^{k}
\end{equation}
In view of the contravariant temporal and spatial components of the stress-energy-momentum tensor~\eqref{stress-energy-momentum 00 bis},~\eqref{stress-energy-momentum j0 bis},~\eqref{stress-energy-momentum 0j bis} and~\eqref{stress-energy-momentum jk bis}, the contravariant components of the stress-energy-momentum tensor are written as,~\cite{Hobson:2006,Schutz:2009}
\begin{equation}\label{stress-energy-momentum mu nu dust}
\boxed{T^{\mu\nu} = \tilde{\rho}\,u^{\mu}\,u^{\nu}}
\end{equation}
In view of the rest energy density~\eqref{energy density rest}, the contravariant components~\eqref{stress-energy-momentum mu nu dust} are recast as,
\begin{equation}\label{stress-energy-momentum mu nu dust second}
\boxed{T^{\mu\nu} = \frac{\tilde{e}}{c^2}\,u^{\mu}\,u^{\nu}}
\end{equation}
%


\section{Dynamics of a relativistic perfect fluid}
\label{Dynamics of a relativistic perfect fluid}

\noindent In order to generalise the relativistic dynamics of a dust described in \S~\ref{Dynamics of a relativistic dust}, we consider in this section a relativistic perfect fluid, which is a continuous medium consisting of relativistic particles undergoing elastic collisions. In particular, we derive the relativistic transformation law for the internal energy density.
For a relativistic dust, the internal energy density vanishes. However, for a relativistic perfect fluid in the local rest frame $\mathcal{\tilde{R}}$, the difference between the energy density $\tilde{e}$ and the internal energy density $\tilde{u}$ is the rest energy density $\tilde{\rho}\,c^2$,
\begin{equation}\label{internal energy density rest}
\tilde{e}-\,\tilde{u} = \tilde{\rho}\,c^2
\end{equation}
For a relativistic perfect fluid in the inertial frame $\mathcal{R}$, the difference between the energy density $e$ and the internal energy density $u$ is the sum of the rest energy density $\tilde{\rho}\,c^2$ and the kinetic energy density $\left(\gamma^2-\,1\right)\tilde{\rho}\,c^2$,
\begin{equation}\label{internal energy density}
e-\,u = \tilde{\rho}\,c^2 + \left(\gamma^2 -\,1\right)\tilde{\rho}\,c^2 = \gamma^2\,\tilde{\rho}\,c^2
\end{equation}
In view of the relativistic transformation law for the mass density~\eqref{density rho frames}, the energy density difference~\eqref{internal energy density} becomes,
\begin{equation}\label{internal energy density bis}
e-\,u = \rho\,c^2
\end{equation}
The energy density difference~\eqref{internal energy density} in the inertial frame $\mathcal{R}$ is related to the energy density difference~\eqref{internal energy density rest} in the local rest frame $\mathcal{\tilde{R}}$ by,
\begin{equation}\label{density e-u frames}
e-\,u = \gamma^2\left(\tilde{e}-\,\tilde{u}\right)
\end{equation}
In view of the relativistic transformation laws for the energy density~\eqref{density e frames} and the energy density difference~\eqref{density e-u frames}, the internal energy density $u$ in the inertial frame $\mathcal{R}$ is related to the internal energy density $\tilde{u}$ in the local rest frame $\mathcal{\tilde{R}}$ by,
\begin{equation}\label{density u frames}
\boxed{u = \gamma^2\,\tilde{u}}
\end{equation}
%


\section{Stress-energy-momentum tensor of a relativistic perfect fluid}
\label{Stress-energy-momentum tensor of a relativistic perfect fluid}

\noindent We determine the contravariant components $T^{\mu\nu}$ of a stress energy-momentum tensor for a relativistic perfect fluid. Such a fluid is characterised by the random motion of the microscopic constituents that generates an isotropic pressure on a macroscopic scale. Thus, in the local rest frame $\mathcal{\tilde{R}}$, the spatial part of the stress-energy-momentum tensor of a relativistic perfect fluid is the identity tensor $\mathbb{1}_3$ multiplied by the scalar pressure $\tilde{p}$ in the local rest frame. It is written in contravariant components as,
\begin{equation}\label{stress-energy-momentum tensor jk LRF}
\tilde{T}^{jk} = \tilde{p}\,\delta^{jk}
\end{equation}
Since a relativistic perfect fluid is a generalisation of a relativistic dust, the internal energy density $\tilde{u}$ needs to be added to the rest energy density $\tilde{\rho}\,c^2$. In order to generalise the contravariant time component of the stress-energy-momentum tensor of a dust~\eqref{stress-energy-momentum tensor 00 LRF}, the contravariant time component of the stress-energy-momentum tensor of a relativistic perfect fluid in the local rest frame $\mathcal{\tilde{R}}$ is written as,
\begin{equation}\label{stress-energy-momentum tensor 00 LRF bis}
\tilde{T}^{00} = \tilde{\rho}\,c^2 + \tilde{u} = \tilde{e}
\end{equation}
In view of the temporal and spatial contravariant components of the tress-energy-momentum tensor~\eqref{stress-energy-momentum tensor jk LRF} and~\eqref{stress-energy-momentum tensor 00 LRF bis}, the contravariant components of the stress-energy-momentum tensor in the local rest frame $\mathcal{\tilde{R}}$ read,
\begin{equation}\label{stress-energy-momentum tensor mu nu fluid LRF}
\tilde{T}^{\mu\nu} = \left(\tilde{\rho}\,c^2 + \tilde{u}\right)\delta^{\mu 0}\,\delta^{\nu 0} + \tilde{p}\left(\eta^{\mu\nu} + \delta^{\mu 0}\,\delta^{\nu 0}\right)
\end{equation}
The contravariant components of the stress-energy-momentum tensor for a perfect fluid~\eqref{stress-energy-momentum tensor mu nu fluid LRF} in the local rest frame $\mathcal{\tilde{R}}$ reduce to the contravariant components of the stress-energy-momentum tensor for relativistic dust~\eqref{stress-energy-momentum tensor mu nu LRF 0} in the absence of pressure and internal energy density, i.e. $\tilde{p} = \tilde{u} = 0$. The contravariant components of the stress-energy-momentum tensor~\eqref{stress-energy-momentum tensor mu nu fluid LRF} in the local rest frame $\mathcal{\tilde{R}}$ can be recast as,
\begin{equation}\label{stress-energy-momentum tensor mu nu fluid LRF bis}
\tilde{T}^{\mu\nu} = \left(\tilde{\rho} + \frac{1}{c^2}\left(\tilde{u} + \tilde{p}\right)\right)c^2\,\delta^{\mu 0}\,\delta^{\nu 0} + \tilde{p}\,\eta^{\mu\nu}
\end{equation}
In relativistic thermodynamics, the inertia density $\tilde{m}$ in the local rest frame $\mathcal{\tilde{R}}$ is different from the mass density $\tilde{\rho}$ since the pressure and the internal energy carry inertia. It is defined as,
\begin{equation}\label{inertia density}
\boxed{\tilde{m} = \tilde{\rho} + \frac{1}{c^2}\left(\tilde{u} + \tilde{p}\right) = \frac{1}{c^2}\left(\tilde{e} + \tilde{p}\right)}
\end{equation}
In view of the inertia density~\eqref{inertia density}, the contravariant components of the stress-energy-momentum tensor~\eqref{stress-energy-momentum tensor mu nu fluid LRF bis} in the local rest frame $\mathcal{\tilde{R}}$ reduce to,
\begin{equation}\label{stress-energy-momentum tensor mu nu fluid LRF quad}
\tilde{T}^{\mu\nu} = \tilde{m}\,c^2\,\delta^{\mu 0}\,\delta^{\nu 0} + \tilde{p}\,\eta^{\mu\nu}
\end{equation}
Using the contravariant components of the velocity vector~\eqref{velocity LRF} in the local rest frame $\mathcal{\tilde{R}}$, the contravariant components of the stress-energy-momentum tensor~\eqref{stress-energy-momentum tensor mu nu fluid LRF quad} in the local rest frame $\mathcal{\tilde{R}}$ reduce to,
\begin{equation}\label{stress-energy-momentum tensor mu nu fluid LRF ter}
\boxed{\tilde{T}^{\mu\nu} = \tilde{m}\,\tilde{u}^{\mu}\,\tilde{u}^{\nu} + \tilde{p}\,\eta^{\mu\nu}}
\end{equation}
In view of the inertia density~\eqref{inertia density}, the contravariant components of the stress-energy-momentum tensor~\eqref{stress-energy-momentum tensor mu nu fluid LRF ter} in the local rest frame $\mathcal{\tilde{R}}$ are recast as,
\begin{equation}\label{stress-energy-momentum tensor mu nu fluid LRF pent}
\boxed{\tilde{T}^{\mu\nu} = \left(\frac{\tilde{e} + \tilde{p}}{c^2}\right)\tilde{u}^{\mu}\,\tilde{u}^{\nu} + \tilde{p}\,\eta^{\mu\nu}}
\end{equation}
The contravariant components of the stress-energy-momentum tensor $T^{\mu\nu}$ of relativistic perfect fluid in the inertial frame $\mathcal{R}$ are obtained by performing inverse Lorentz transformations on the contravariant components of the stress-energy-momentum tensor $\tilde{T}^{\mu\nu}$ of a relativistic perfect fluid in in the local rest frame $\mathcal{\tilde{R}}$,
\begin{equation}\label{stress-energy-momentum tensor mu nu Lorentz fluid}
T^{\mu\nu} = \left(\Lambda^{-1}\right)^\mu_{\phantom{\mu}\rho}\left(\Lambda^{-1}\right)^\nu_{\phantom{\nu}\lambda}\,\tilde{T}^{\rho\lambda}
\end{equation}
Using the components of the stress-energy-momentum tensor~\eqref{stress-energy-momentum tensor mu nu fluid LRF quad} in the local rest frame $\mathcal{\tilde{R}}$, the components of the stress-energy-momentum tensor~\eqref{stress-energy-momentum tensor mu nu Lorentz fluid} in the inertial frame $\mathcal{R}$ are recast as,
\begin{equation}\label{stress-energy-momentum tensor mu nu Lorentz bis fluid}
\begin{split}
&T^{\mu\nu} = \left(\Lambda^{-1}\right)^\mu_{\phantom{\mu}0}\left(\Lambda^{-1}\right)^\nu_{\phantom{0}0}\,\tilde{m}\,c^2\\
&\phantom{T^{\mu\nu} =} + \left(\left(\Lambda^{-1}\right)^{\mu}_{\phantom{0}0}\left(\Lambda^{-1}\right)^{\nu}_{\phantom{0}0}\eta^{00} + \left(\Lambda^{-1}\right)^{\mu}_{\phantom{0}j}\left(\Lambda^{-1}\right)^{\nu}_{\phantom{0}k}\eta^{jk}\right)\tilde{p}
\end{split}
\end{equation}
The contravariant time component of the stress-energy-momentum tensor~\eqref{stress-energy-momentum tensor mu nu Lorentz bis fluid} of a relativistic fluid in the inertial frame $\mathcal{R}$ reads,
\begin{equation}\label{stress-energy-momentum fluid 00}
\begin{split}
&T^{00} = \left(\Lambda^{-1}\right)^0_{\phantom{0}0}\left(\Lambda^{-1}\right)^0_{\phantom{0}0}\,\tilde{m}\,c^2\\
&\phantom{T^{00} =} + \left(\left(\Lambda^{-1}\right)^0_{\phantom{0}0}\left(\Lambda^{-1}\right)^0_{\phantom{0}0}\eta^{00} + \left(\Lambda^{-1}\right)^0_{\phantom{0}j}\left(\Lambda^{-1}\right)^0_{\phantom{0}k}\eta^{jk}\right)\tilde{p}
\end{split}
\end{equation}
Using the Minkowski metric~\eqref{metric}, the Lorentz factor~\eqref{Lorentz factor v} and the components of the inverse Lorentz transformation~\eqref{Lorentz transformation inverse components}, we obtain the relation,
\begin{equation}\label{velocity jk}
v^{j}\,v^{k} = \left(v^{i}\,\delta^{j}_{\phantom{j}i}\right)\left(v^{i}\,\delta^{k}_{\phantom{k}i}\right) = v^{i\,2}\,\delta^{jk} = v^2\,\delta^{jk}
\end{equation}
which yields the identity,
\begin{equation}\label{tensor rel 00 bis}
\left(\Lambda^{-1}\right)^0_{\phantom{0}0}\left(\Lambda^{-1}\right)^0_{\phantom{0}0}\eta^{00} + \left(\Lambda^{-1}\right)^0_{\phantom{0}j}\left(\Lambda^{-1}\right)^0_{\phantom{0}k}\eta^{jk} = \gamma^2\left(\frac{v^2}{c^2}-\,1\right) = -\,1
\end{equation}
Taking into account the metric~\eqref{metric}, the contravariant time component of the velocity vector~\eqref{temporal velocity contravariant} and the identity~\eqref{tensor rel 00 bis}, the contravariant time component of the stress-energy-momentum tensor of a relativistic perfect fluid~\eqref{stress-energy-momentum fluid 00} reduces to,
\begin{equation}\label{stress-energy-momentum fluid 00 bis}
T^{00} = \tilde{m}\left(\gamma\,c\right)\left(\gamma\,c\right) -\,\tilde{p} = \tilde{m}\,u^{0}\,u^{0} + \tilde{p}\,\eta^{00}
\end{equation}
The contravariant space and time components of the stress-energy-momentum tensor~\eqref{stress-energy-momentum tensor mu nu Lorentz bis fluid} of a relativistic fluid in the inertial frame $\mathcal{R}$ read,
\begin{equation}\label{stress-energy-momentum fluid j0}
\begin{split}
&T^{j0} = \left(\Lambda^{-1}\right)^j_{\phantom{j}0}\left(\Lambda^{-1}\right)^0_{\phantom{0}0}\,\tilde{m}\,c^2\\
&\phantom{T^{j0} =} + \left(\left(\Lambda^{-1}\right)^j_{\phantom{j}0}\left(\Lambda^{-1}\right)^0_{\phantom{0}0}\eta^{00} + \left(\Lambda^{-1}\right)^j_{\phantom{j}k}\left(\Lambda^{-1}\right)^0_{\phantom{0}\ell}\eta^{k\ell}\right)\tilde{p}
\end{split}
\end{equation}
Using the Minkowski metric~\eqref{metric} and the components of the inverse Lorentz transformation~\eqref{Lorentz transformation inverse components}, we obtain the identity,
\begin{equation}\label{tensor rel j0 bis}
\left(\Lambda^{-1}\right)^j_{\phantom{j}0}\left(\Lambda^{-1}\right)^0_{\phantom{0}0}\eta^{00} + \left(\Lambda^{-1}\right)^j_{\phantom{j}k}\left(\Lambda^{-1}\right)^0_{\phantom{0}\ell}\eta^{k\ell} = -\,\frac{\gamma^2}{c}\,\left(v^{j} -\,\delta^{j}_{\phantom{j}k}\,v^{k}\right) = 0
\end{equation}
Taking into account the metric~\eqref{metric}, the contravariant temporal and spatial components of the velocity vector~\eqref{temporal velocity contravariant} and~\eqref{spatial velocity contravariant and covariant}, and the identity~\eqref{tensor rel j0 bis}, the contravariant temporal and spatial components of the stress-energy-momentum tensor of a relativistic perfect fluid~\eqref{stress-energy-momentum fluid j0} reduce to,
\begin{equation}\label{stress-energy-momentum fluid j0 bis}
T^{j0} = \tilde{m}\left(\gamma\,v^{j}\right)\left(\gamma\,c\right) = \tilde{m}\,u^{j}\,u^{0} + \tilde{p}\,\eta^{j0}
\end{equation}
The contravariant space and time components of the stress-energy-momentum tensor~\eqref{stress-energy-momentum tensor mu nu Lorentz bis fluid} of a relativistic fluid in the inertial frame $\mathcal{R}$ read,
\begin{equation}\label{stress-energy-momentum fluid 0j}
\begin{split}
&T^{0j} = \left(\Lambda^{-1}\right)^0_{\phantom{0}0}\left(\Lambda^{-1}\right)^j_{\phantom{j}0}\,\tilde{m}\,c^2\\
&\phantom{T^{0j} =} + \left(\left(\Lambda^{-1}\right)^0_{\phantom{0}0}\left(\Lambda^{-1}\right)^j_{\phantom{j}0}\,\eta^{00} + \left(\Lambda^{-1}\right)^0_{\phantom{0}k}\left(\Lambda^{-1}\right)^j_{\phantom{j}\ell}\,\eta^{k\ell}\right)\tilde{p}
\end{split}
\end{equation}
Using the Minkowski metric~\eqref{metric} and the components of the inverse Lorentz transformation~\eqref{Lorentz transformation inverse components}, we obtain the identity,
\begin{equation}\label{tensor rel 0j bis}
\left(\Lambda^{-1}\right)^0_{\phantom{0}0}\left(\Lambda^{-1}\right)^j_{\phantom{j}0}\,\eta^{00} + \left(\Lambda^{-1}\right)^0_{\phantom{0}k}\left(\Lambda^{-1}\right)^j_{\phantom{j}\ell}\,\eta^{k\ell} = -\,\frac{\gamma^2}{c}\,\left(v^{j} -\,\delta^{j}_{\phantom{j}k}\,v^{k}\right) = 0
\end{equation}
Taking into account the metric~\eqref{metric}, the contravariant temporal and spatial components of the velocity vector~\eqref{temporal velocity contravariant} and~\eqref{spatial velocity contravariant and covariant}, and the identity~\eqref{tensor rel 0j bis}, the contravariant temporal and spatial components of the stress-energy-momentum tensor of a relativistic perfect fluid~\eqref{stress-energy-momentum fluid 0j} reduce to,
\begin{equation}\label{stress-energy-momentum fluid 0j bis}
T^{0j} = \tilde{m}\left(\gamma\,c\right)\left(\gamma\,v^{j}\right) = \tilde{m}\,u^{j}\,u^{0} + \tilde{p}\,\eta^{0j}
\end{equation}
The contravariant spatial components of the stress-energy-momentum tensor~\eqref{stress-energy-momentum tensor mu nu Lorentz bis fluid} of a relativistic fluid in the inertial frame $\mathcal{R}$ read,
\begin{equation}\label{stress-energy-momentum fluid jk}
\begin{split}
&T^{jk} = \left(\Lambda^{-1}\right)^j_{\phantom{j}0}\left(\Lambda^{-1}\right)^k_{\phantom{k}0}\,\tilde{m}\,c^2\\
&\phantom{T^{jk} =} + \left(\left(\Lambda^{-1}\right)^j_{\phantom{j}0}\left(\Lambda^{-1}\right)^k_{\phantom{k}0}\,\eta^{00} + \left(\Lambda^{-1}\right)^j_{\phantom{j}\ell}\left(\Lambda^{-1}\right)^k_{\phantom{k}m}\,\eta^{\ell m}\right)\tilde{p}
\end{split}
\end{equation}
Using the Minkowski metric~\eqref{metric}, the components of the inverse Lorentz transformation~\eqref{Lorentz transformation inverse components} and the relation~\eqref{velocity jk}, we obtain the identity,
\begin{equation}\label{tensor rel jk bis}
\left(\Lambda^{-1}\right)^j_{\phantom{j}0}\left(\Lambda^{-1}\right)^k_{\phantom{k}0}\,\eta^{00} + \left(\Lambda^{-1}\right)^j_{\phantom{j}\ell}\left(\Lambda^{-1}\right)^k_{\phantom{k}m}\,\eta^{\ell m} = -\,\gamma^2\left(\frac{v^2}{c^2} -\,1\right)\delta^{jk} = \delta^{jk}
\end{equation}
Taking into account the metric~\eqref{metric}, the contravariant spatial components of the velocity vector~\eqref{spatial velocity contravariant and covariant} and the identity~\eqref{tensor rel jk bis}, the contravariant spatial components of the stress-energy-momentum tensor of a relativistic perfect fluid~\eqref{stress-energy-momentum fluid jk} reduce to,
\begin{equation}\label{stress-energy-momentum fluid jk bis}
T^{jk} = \tilde{m}\left(\gamma\,v^{j}\right)\left(\gamma\,v^{k}\right) + \tilde{p}\,\delta^{jk} = \tilde{m}\,u^{j}\,u^{k} + \tilde{p}\,\eta^{jk}
\end{equation}
In view of the contravariant temporal and spatial components of the stress-energy-momentum tensor of a relativistic perfect fluid~\eqref{stress-energy-momentum fluid 00 bis},~\eqref{stress-energy-momentum fluid j0 bis},~\eqref{stress-energy-momentum fluid 0j bis} and~\eqref{stress-energy-momentum fluid jk bis}, the contravariant components of the stress-energy-momentum tensor of a relativistic perfect fluid in the inertial frame $\mathcal{R}$ are written as,
\begin{equation}\label{stress-energy-momentum tensor fluid mu nu}
\boxed{T^{\mu\nu} = \tilde{m}\,u^{\mu}\,u^{\nu} + \tilde{p}\,\eta^{\mu\nu}}
\end{equation}
In view of the inertia density~\eqref{inertia density}, the contravariant components of the stress-energy-momentum tensor of a relativistic perfect fluid~\eqref{stress-energy-momentum tensor fluid mu nu} are recast as,~\cite{Hobson:2006,Schutz:2009}
\begin{equation}\label{stress-energy-momentum tensor fluid mu nu bis}
\boxed{T^{\mu\nu} = \left(\frac{\tilde{e} + \tilde{p}}{c^2}\right)u^{\mu}\,u^{\nu} + \tilde{p}\,\eta^{\mu\nu}}
\end{equation}
%


\section{Relativistic temperature}
\label{Relativistic temperature}

\noindent We derive the relativistic transformation law for the temperature in this section, which sheds light on the Einstein, Planck and Ott controversy.~\cite{Einstein:1907,Planck:1908,Ott:1963} In the local rest frame $\mathcal{\tilde{R}}$, the temperature $\tilde{T}$ is defined as,
\begin{equation}\label{temperature LRF}
\tilde{T} = \frac{\partial \tilde{u}}{\partial \tilde{s}}
\end{equation}
In the inertial frame $\mathcal{R}$, the temperature $T$ is defined as,
\begin{equation}\label{temperature inertial}
T = \frac{\partial u}{\partial s}
\end{equation}
In view of the transformation laws for the internal energy density~\eqref{density u frames} and the entropy density~\eqref{density s frames}, the temperature in the inertial frame $\mathcal{R}$ is recast as,
\begin{equation}\label{temperature inertial bis}
T = \frac{\partial\left(\gamma^2\,\tilde{u}\right)}{\partial\left(\gamma\,\tilde{s}\right)} = \frac{\partial\left(\gamma^2\,\tilde{u}\right)}{\partial\,\tilde{s}}\left(\frac{\partial\left(\gamma\,\tilde{s}\right)}{\partial\,\tilde{s}}\right)^{-1}
\end{equation}
Since the internal energy density $\tilde{u}$ and the entropy density $\tilde{s}$ in the local rest frame $\mathcal{\tilde{R}}$ are independent of the Lorentz factor $\gamma$, we obtain the identities, 
\begin{equation}\label{u and s independent gamma}
\frac{\partial\left(\gamma^2\,\tilde{u}\right)}{\partial\,\tilde{s}} = \gamma^2\,\frac{\partial\,\tilde{u}}{\partial\,\tilde{s}} \qquad\textrm{and}\qquad \frac{\partial\left(\gamma\,\tilde{s}\right)}{\partial\,\tilde{s}} = \gamma
\end{equation}
According to the identities~\eqref{u and s independent gamma}, the temperature~\eqref{temperature inertial bis} in the inertial frame $\mathcal{R}$ reduces to, 
\begin{equation}\label{temperature inertial ter}
T = \gamma\,\frac{\partial \tilde{u}}{\partial \tilde{s}}
\end{equation}
The temperature~\eqref{temperature inertial ter} in the inertial frame $\mathcal{R}$ is related to the temperature~\eqref{temperature LRF} in the local rest frame $\mathcal{\tilde{R}}$ by,
\begin{equation}\label{temperature frames}
\boxed{T = \gamma\,\tilde{T}}
\end{equation}
which implies that a relativistic fluid appears hotter when it is in relativistic motion. This is the conclusion reached by Einstein in a private letter to von Laue~\cite{Liu:1992} and by Ott~\cite{Ott:1963,Farias:2017} in the famous controversy that opposed their results to the previous work performed by Einstein and Planck.~\cite{Einstein:1907,Planck:1908} In fact, the initial conclusion of Einstein and Planck can be reached by defining the temperature $T^{\prime}$ in the inertial frame $\mathcal{R}$ as,
\begin{equation}\label{temperature planck}
T^{\prime} = \frac{\partial \tilde{u}}{\partial s} = \frac{\partial \tilde{u}}{\partial\left(\gamma\,\tilde{s}\right)} = \frac{\partial \tilde{u}}{\partial \tilde{s}}\left(\frac{\partial\left(\gamma\,\tilde{s}\right)}{\partial\,\tilde{s}}\right)^{-1}
\end{equation}
According to identities~\eqref{u and s independent gamma}, the temperature~\eqref{temperature planck} reduces to, 
\begin{equation}\label{temperature planck bis}
T^{\prime} = \frac{1}{\gamma}\,\frac{\partial \tilde{u}}{\partial \tilde{s}}
\end{equation}
The temperature~\eqref{temperature planck bis} in the inertial frame $\mathcal{R}$ is related to the temperature~\eqref{temperature LRF} in the local rest frame $\mathcal{\tilde{R}}$ by,
\begin{equation}\label{temperature planck relation}
T^{\prime} = \frac{1}{\gamma}\,\tilde{T}
\end{equation}
The relativistic transformation~\eqref{temperature planck relation} for the temperature is flawed for the following reason : the internal energy density is not frame independent in a relativistic framework according relation~\eqref{density u frames}. Therefore, the temperature in the inertial frame $\mathcal{R}$ has to be defined as~\eqref{temperature inertial ter} rather than~\eqref{temperature planck bis}. The factor $\gamma^2$ associated with the Lorentz transformation of the internal energy density is essential in deriving this result. In order to test the relativistic transformation law for the temperature, we can do the following ``Gedankenexperiment'' : a black body is moving at relativistic velocity with respect to the inertial frame $\mathcal{R}$ of the observer. A photon of frequency $\tilde{\nu}$ in the local rest frame $\mathcal{\tilde{R}}$ of the black body is emitted orthogonally to the direction of motion of the black body. This photon is detected in the inertial frame $\mathcal{R}$ of the observer with a frequency $\nu$. Since the photon is emitted orthogonally to the direction of motion of motion, the detected frequency $\nu$ is related to the emitted frequency $\tilde{\nu}$ through a relativistic transverse Doppler effect,
\begin{equation}\label{Doppler effect frequency}
\nu = \gamma\,\tilde{\nu}
\end{equation}
In the local rest frame $\mathcal{\tilde{R}}$ of the black body, the emitted spectral radiance of the black body is a function of the emitted frequency $\tilde{\nu}$ of the photon and the temperature $\tilde{T}$ given by Planck's law,
\begin{equation}\label{spectral radiance Planck law BB}
\tilde{B}\left(\tilde{\nu},\tilde{T}\right) = \frac{2h\tilde{\nu}^3}{c^2}\,\frac{1}{\displaystyle{e^{\frac{h\tilde{\nu}}{k_B\tilde{T}}}}-\,1}
\end{equation}
In the inertial frame $\mathcal{R}$ of the observer, the detected spectral radiance of the black body is a function of the detected frequency $\nu$ of the photon and the measured temperature $T$ given by Planck's law,
\begin{equation}\label{spectral radiance Planck law observer}
B\left(\nu,T\right) = \frac{2h\nu^3}{c^2}\,\frac{1}{\displaystyle{e^{\frac{h\nu}{k_BT}}}-\,1}
\end{equation}
As shown by Johnson and Teller~\cite{Johnson:1982}, the ratio of the spectral radiance and the frequency cubed is frame invariant,
\begin{equation}\label{Johnson and Teller}
\frac{\tilde{B}\left(\tilde{\nu},\tilde{T}\right)}{\tilde{\nu}^3} = \frac{B\left(\nu,T\right)}{\nu^3}
\end{equation}
which implies that,
\begin{equation}\label{invariance Doppler}
\frac{h\tilde{\nu}}{k_B\tilde{T}} = \frac{h\nu}{k_BT}
\end{equation}
In view of the frame invariance~\eqref{invariance Doppler} and the transverse Doppler effect~\eqref{Doppler effect frequency}, the temperature $T$ of the black body measured in the inertial frame $\mathcal{R}$ of the observer is related to the temperature of the black body $\tilde{T}$ in the local rest frame $\mathcal{\tilde{R}}$ of the black body by,
\begin{equation}\label{Doppler effect temperature}
\frac{T}{\tilde{T}} = \frac{\nu}{\tilde{\nu}} = \gamma \qquad\text{thus}\qquad T = \gamma\,\tilde{T}
\end{equation}
in agreement with the prediction~\eqref{temperature frames} as it should.
%


\section{Relativistic chemical potential}
\label{Relativistic chemical potential}

\noindent We derive the relativistic transformation law for the chemical potential in this section. In the local rest frame $\mathcal{\tilde{R}}$, the chemical potential $\tilde{\mu}$ is defined as,
\begin{equation}\label{chemical potential LRF}
\tilde{\mu} = \frac{\partial \tilde{u}}{\partial \tilde{n}}
\end{equation}
In the inertial frame $\mathcal{R}$, the chemical potential $\mu$ is defined as,
\begin{equation}\label{chemical potential inertial}
\mu = \frac{\partial u}{\partial n}
\end{equation}
In view of the transformation laws for the internal energy density~\eqref{density u frames} and the number density density~\eqref{density n frames}, the chemical potential in the inertial frame $\mathcal{R}$ is recast as,
\begin{equation}\label{chemical potential inertial bis}
\mu = \frac{\partial\left(\gamma^2\,\tilde{u}\right)}{\partial\left(\gamma\,\tilde{n}\right)} = \frac{\partial\left(\gamma^2\,\tilde{u}\right)}{\partial\,\tilde{n}}\left(\frac{\partial\left(\gamma\,\tilde{n}\right)}{\partial\,\tilde{n}}\right)^{-1}
\end{equation}
Since the internal energy density $\tilde{u}$ and the number density $\tilde{n}$ in the local rest frame $\mathcal{\tilde{R}}$ are independent of the Lorentz factor $\gamma$, we obtain the identities, 
\begin{equation}\label{u and n independent gamma}
\frac{\partial\left(\gamma^2\,\tilde{u}\right)}{\partial\,\tilde{n}} = \gamma^2\,\frac{\partial\,\tilde{u}}{\partial\,\tilde{n}} \qquad\textrm{and}\qquad \frac{\partial\left(\gamma\,\tilde{n}\right)}{\partial\,\tilde{n}} = \gamma
\end{equation}
According to the identities~\eqref{u and n independent gamma}, the chemical potential~\eqref{chemical potential inertial bis} in the inertial frame $\mathcal{R}$ reduces to,
\begin{equation}\label{chemical potential inertial ter}
\mu = \gamma\,\frac{\partial \tilde{u}}{\partial \tilde{n}}
\end{equation}
The chemical potential~\eqref{chemical potential inertial ter} in the inertial frame $\mathcal{R}$ is related to the chemical potential~\eqref{chemical potential LRF} in the local rest frame $\mathcal{\tilde{R}}$ by,~\cite{Przanowski:2011}
\begin{equation}\label{chemical potential frames}
\boxed{\mu = \gamma\,\tilde{\mu}}
\end{equation}
which implies that a relativistic fluid has a higher chemical potential when it is in relativistic motion.


\section{Relativistic pressure}
\label{Relativistic pressure}

\noindent We use the relativistic transformation laws for the internal energy, the entropy, the number density, the temperature and the chemical potential in order to derive the relativistic transformation law for the pressure in this section. In order to do so, we begin with the Euler relation per unit volume in the local rest frame $\mathcal{\tilde{R}}$,
\begin{equation}\label{Euler relation per unit volume rest}
\tilde{u} = \tilde{T}\,\tilde{s} -\,\tilde{p} -\,\tilde{\mu}\,\tilde{n}
\end{equation}
which implies that the pressure in the local rest frame $\mathcal{\tilde{R}}$ is given by,
\begin{equation}\label{pressure rest}
\tilde{p} = \tilde{T}\,\tilde{s} -\,\tilde{\mu}\,\tilde{n} -\,\tilde{u}
\end{equation}
The Euler relation per unit volume in the inertial frame $\mathcal{R}$ reads,
\begin{equation}\label{Euler relation per unit volume inertial}
u = T\,s -\,p -\,\mu\,n
\end{equation}
which implies that the pressure in the inertial frame $\mathcal{R}$ is given by,
\begin{equation}\label{pressure inertial}
p = T\,s -\,\mu\,n -\,u
\end{equation}
In view of the transformation laws for the entropy density~\eqref{density s frames}, the number density~\eqref{density n frames}, the internal energy density~\eqref{density u frames}, the temperature~\eqref{temperature frames} and the chemical potential~\eqref{chemical potential frames}, the pressure~\eqref{pressure inertial} in the inertial frame $\mathcal{R}$ is recast as,
\begin{equation}\label{pressure inertial bis}
p = (\gamma\,\tilde{T})\left(\gamma\,\tilde{s}\right) -\,\left(\gamma\,\tilde{\mu}\right)\left(\gamma\,\tilde{n}\right) -\,(\gamma^2\,\tilde{u})
\end{equation}
Thus, in agreement with Sutcliffe~\cite{Sutcliffe:1965}, the pressure~\eqref{pressure inertial bis} in the inertial frame $\mathcal{R}$ is related to the pressure~\eqref{pressure rest} in the local rest frame $\mathcal{\tilde{R}}$ by,
\begin{equation}\label{pressure frames}
\boxed{p = \gamma^2\,\tilde{p}}
\end{equation}
which implies that a relativistic fluid has a larger pressure when it is in relativistic motion. The thermodynamic pressure that satisfies the transformation law~\eqref{pressure frames} has to be distinguished from the mechanical pressure that is invariant under a relativistic frame transformation, as noted by Einstein and Planck~\cite{Einstein:1907,Planck:1908}. As mentioned by Sutcliffe : ``The two pressures are the same only in the local rest frame $\mathcal{\tilde{R}}$. This distinction is necessary in order for relativistic thermodynamics to be consistent with relativistic mechanics.''~\cite{Sutcliffe:1965} In the non-relativistic limit, the thermodynamic and mechanical pressures coincide. As an illustration, the equation of state of an ideal gas in the local rest frame $\mathcal{\tilde{R}}$ reads,~\cite{Callen:1985}
\begin{equation}\label{ideal gas rest}
\tilde{p} = \tilde{n}\,R\,\tilde{T} \qquad\textrm{(ideal gas)}
\end{equation}
where $R$ is the ideal gas constant. Multiplying the equation of state~\eqref{ideal gas rest} in the local rest frame $\mathcal{\tilde{R}}$ by $\gamma^2$ yields,
\begin{equation}\label{ideal gas rest bis}
\left(\gamma^2\,\tilde{p}\right) = \left(\gamma\,\tilde{n}\right)R\,(\gamma\,\tilde{T})
\end{equation}
Using the transformation laws for the number density~\eqref{density n frames}, the temperature~\eqref{temperature frames} and the pressure~\eqref{pressure frames}, equation~\eqref{ideal gas rest bis} is identified as the equation of state of an ideal gas in the inertial frame $\mathcal{R}$,~\cite{Callen:1971}
\begin{equation}\label{ideal gas inertial}
p = n\,R\,T \qquad\textrm{(ideal gas)}
\end{equation}
as expected. In order to test the relativistic transformation law for the pressure, we can do the following ``Gedankenexperiment'' : a black body is moving at relativistic velocity with respect to the inertial frame $\mathcal{R}$ of the observer. A monochromatic filter selects the photons of frequency $\tilde{\nu}$ and density $\tilde{n}$ in the local rest frame $\mathcal{\tilde{R}}$ of the black body that are emitted orthogonally to the direction of motion of the black body. The photons are detected in the inertial frame $\mathcal{R}$ of the observer with a frequency $\nu$ and a density $n$. Since the photons are emitted orthogonally to the direction of motion of motion, the detected frequency $\nu$ is related to the emitted frequency $\tilde{\nu}$ through the relativistic transverse Doppler effect~\eqref{Doppler effect frequency}. The internal energy density $\tilde{u}$ of the emitted monochromatic photons in the local rest frame $\mathcal{\tilde{R}}$ of the black body is given by the Planck-Einstein relation,
\begin{equation}\label{Planck-Einstein BB}
\tilde{u} = \tilde{n}\,\hbar\,\tilde{\nu}
\end{equation}
The internal energy density $u$ of the detected monochromatic photons in the local rest frame $\mathcal{R}$ of the observer is given by the Planck-Einstein relation,
\begin{equation}\label{Planck-Einstein observer}
u = n\,\hbar\,\nu
\end{equation}
The radiation pressure $\tilde{p}$ of the emitted photons in the local rest frame $\mathcal{\tilde{R}}$ of the black body reads,
\begin{equation}\label{radiation pressure BB}
\tilde{p} = \frac{1}{3}\,\tilde{u} = \frac{1}{3}\,\tilde{n}\,\hbar\,\tilde{\nu}
\end{equation}
The radiation pressure $p$ of the detected photons in the inertial frame $\mathcal{R}$ of the observer reads,
\begin{equation}\label{radiation pressure observer}
p = \frac{1}{3}\,u = \frac{1}{3}\,n\,\hbar\,\nu
\end{equation}
In view of the transformation law~\eqref{density n frames} for the number density and the relativistic transverse Doppler effect~\eqref{Doppler effect frequency}, the radiation pressure $p$ of the detected photons~\eqref{radiation pressure observer} in the inertial frame $\mathcal{R}$ of the observer is related to the radiation pressure $\tilde{p}$ of the emitted photons~\eqref{radiation pressure BB} in the local rest frame $\mathcal{\tilde{R}}$ of the black body by the transformation law,
\begin{equation}\label{radiation pressure transformnation law}
p = \frac{1}{3}\,n\,\hbar\,\nu = \frac{1}{3}\left(\gamma\,\tilde{n}\right)\hbar\left(\gamma\,\tilde{\nu}\right) = \gamma^2\,\tilde{p}
\end{equation}
in agreement with the prediction~\eqref{pressure frames} as it should.


\section{Relativistic force density}
\label{Relativistic force density}

\noindent We determine the properties of the contravariant components of the force density vector in this section. According to the relativistic equation of motion~\eqref{relativistic law of motion} of a particle, the contravariant components of the force $F^{\mu}$ exerted on an infinitesimal fluid element are defined as the derivative of the contravariant components of the momentum $P^{\mu}$ with respect to the proper time $\tilde{t}$ in the local rest frame $\mathcal{\tilde{R}}$. Thus, the contravariant components of the force vector $F^{\mu}$ exerted on an infinitesimal fluid element are obtained by multiplying the contravariant components of the force density vector $f^{\mu}$ by the infinitesimal volume $d\tilde{V}_3$ in the local rest frame $\mathcal{\tilde{R}}$,~\cite{Mihalas:1984}
\begin{equation}\label{force density contravariant}
F^{\mu} = f^{\mu}\,d\tilde{V}_3
\end{equation}
In view of the contravariant components of the force vector~\eqref{force density contravariant}, the contravariant time component of the force vector $F^{0}$ is related to the contravariant time component of the force density vector $f^{0}$ by,
\begin{equation}\label{force density time}
F^{0} = f^{0}\,d\tilde{V}_3
\end{equation}
and the contravariant spatial component of the force vector $F^{j}$ is related to the contravariant spatial component of the force density vector $f^{j}$ by,
\begin{equation}\label{force density space}
F^{j} = f^{j}\,d\tilde{V}_3 
\end{equation}
The spatial force density vector $\boldsymbol{f}$ in the inertial frame $\mathcal{R}$ is defined as,
\begin{equation}\label{force density spatial 0}
\boldsymbol{f} = f^{j}\,\boldsymbol{e}_j
\end{equation}
In view of the volume contraction~\eqref{volume contraction}, the spatial force vector~\eqref{force}, the spatial contravariant component of the force vector~\eqref{force density space} and the spatial force density vector~\eqref{force density spatial 0}, the force vector is expressed as,
\begin{equation}\label{force density spatial}
\boldsymbol{F} = \frac{1}{\gamma}\,F^{j}\,\boldsymbol{e}_j = \frac{1}{\gamma}\,f^{j}\,\boldsymbol{e}_j\,d\tilde{V}_3 = f^{j}\,\boldsymbol{e}_j\,dV_3 = \boldsymbol{f}\,dV_3 
\end{equation}
According to the contravariant time component of the force vector~\eqref{force density time} and the spatial force vector~\eqref{force density spatial}, the contravariant time component of the force vector~\eqref{temporal force contravariant} is recast as,
\begin{equation}\label{temporal force density contravariant 0}
f^{0}\,d\tilde{V}_3 = \frac{\gamma}{c}\,\boldsymbol{f}\cdot\boldsymbol{v}\,dV_3 
\end{equation}
In view of the volume contraction~\eqref{volume contraction} and the contravariant time component of the force vector~\eqref{temporal force density contravariant 0}, the power density in the inertial frame $\mathcal{R}$ is given by,~\cite{Mihalas:1984}
\begin{equation}\label{force density time bis}
\boxed{f^{0} = \frac{1}{c}\,\boldsymbol{f}\cdot\boldsymbol{v}}
\end{equation}
%


\section{Relativistic energy continuity equation}
\label{Relativistic energy continuity equation}

\noindent We determine the relativistic energy continuity equation by taking the contravariant time components of the energy-momentum continuity equation~\eqref{relativistic momentum continuity equation} in this section. In view of the stress-energy-momentum tensor of a relativistic perfect fluid~\eqref{stress-energy-momentum tensor fluid mu nu bis}, the relativistic energy-momentum continuity equation~\eqref{relativistic momentum continuity equation} of a relativistic perfect fluid becomes,~\cite{Rezzolla:2013}
\begin{equation}\label{relativistic momentum continuity equation fluid}
\partial_{\mu}\left(\left(\frac{\tilde{e} + \tilde{p}}{c^2}\right)u^{\mu}\,u^{\nu} + \tilde{p}\,\eta^{\mu\nu}\right) = f^{\nu}
\end{equation}
In view of the Minkowski metric~\eqref{metric}, the contravariant components of the velocity vector~\eqref{temporal velocity contravariant} and the contravariant time component of the force density~\eqref{force density time bis}, the component $\nu = 0$ of the relativistic energy-momentum continuity equation~\eqref{relativistic momentum continuity equation fluid} yields the relativistic energy continuity equation,
\begin{equation}\label{relativistic momentum continuity equation fluid 0}
\partial_{\mu}\left(\gamma\left(\frac{\tilde{e} + \tilde{p}}{c}\right)u^{\mu} -\,\tilde{p}\,\delta^{\mu 0}\right) = \frac{1}{c}\,\boldsymbol{f}\cdot\boldsymbol{v}
\end{equation}
Using the frame duality~\eqref{duality space}, the contravariant temporal and spatial components of the velocity vector~\eqref{temporal velocity contravariant} and~\eqref{spatial velocity contravariant and covariant}, as well as the partial time derivative~\eqref{temporal partial derivative contravariant}, the relativistic energy continuity equation~\eqref{relativistic momentum continuity equation fluid 0} is recast as,
\begin{equation}\label{relativistic momentum continuity equation fluid 1}
\frac{1}{c}\,\partial_{t}\,\Big(\gamma^2\,\tilde{e} + \tilde{p}\left(\gamma^2 -\,1\right)\Big) + \frac{1}{c}\left(\boldsymbol{e}^j\,\partial_{j}\right)\cdot\left(\gamma^2\left(\tilde{e} + \tilde{p}\right)v^{i}\,\boldsymbol{e}_i\right)
= \frac{1}{c}\,\boldsymbol{f}\cdot\boldsymbol{v}
\end{equation}
In view of the spatial velocity vector~\eqref{spatial velocity}, the gradient~\eqref{gradient}, the transformation laws for the energy density~\eqref{density e frames} and the pressure~\eqref{pressure frames}, the relativistic energy continuity equation~\eqref{relativistic momentum continuity equation fluid 1} multiplied by $c$ becomes,
\begin{equation}\label{relativistic momentum continuity equation fluid 2}
\partial_{t}\left(e + \left(1-\,\frac{1}{\gamma^2}\right)p\right) + \boldsymbol{\nabla}\cdot\Big(\left(e + p\right)\boldsymbol{v}\Big) = \boldsymbol{f}\cdot\boldsymbol{v}
\end{equation}
According to the Lorentz factor~\eqref{Lorentz factor v}, the relativistic energy continuity equation~\eqref{relativistic momentum continuity equation fluid 2} in the inertial frame $\mathcal{R}$ is recast as,
\begin{equation}\label{relativistic energy continuity equation fluid}
\boxed{\partial_{t}\left(e + \frac{v^2}{c^2}\,p\right) + \boldsymbol{\nabla}\cdot\left(e\,\boldsymbol{v}\right) = \left(\boldsymbol{f} -\,\boldsymbol{\nabla}\,p\right)\cdot\boldsymbol{v} \qquad\textrm{(relativistic)}}
\end{equation}
where the term,
\begin{equation}\label{relativistic correction}
\frac{v^2}{c^2}\,p =  \left(1-\,\frac{1}{\gamma^2}\right)p = p -\,\tilde{p}
\end{equation}
is a relativistic correction of the energy density due to the pressure difference between the inertial frame $\mathcal{R}$ and the local rest frame $\mathcal{\tilde{R}}$ which leads to a relativistic coupling between heat and work. In the non-relativistic limit, the energy continuity equation~\eqref{relativistic energy continuity equation fluid} reduces to,~\cite{Brechet:2019}
\begin{equation}\label{non relativistic energy continuity equation}
\boxed{\partial_{t}\,e + \boldsymbol{\nabla}\cdot\left(e\,\boldsymbol{v}\right) = \left(\boldsymbol{f} -\,\boldsymbol{\nabla}\,p\right)\cdot\boldsymbol{v} \qquad\textrm{(non-relativistic)}}
\end{equation}
as expected.
%


\section{Relativistic momentum continuity equation}
\label{Relativistic momentum continuity equation}

\noindent We determine the relativistic momentum continuity equation by multiplying the contravariant spatial components of the energy-momentum continuity equation~\eqref{relativistic momentum continuity equation} with the corresponding spatial orthonormal basis vectors in this section. In view of the contravariant spatial components of the velocity vector~\eqref{spatial velocity contravariant and covariant}, the component $\nu = i$ of the relativistic energy-momentum continuity equation~\eqref{relativistic momentum continuity equation fluid} yields the relativistic momentum continuity equation,~\cite{Rezzolla:2013}
\begin{equation}\label{relativistic momentum continuity equation fluid i 0}
\partial_{\mu}\left(\gamma\left(\frac{\tilde{e} + \tilde{p}}{c^2}\right)u^{\mu}v^{i} -\,\tilde{p}\,\delta^{\mu i}\right) = f^{i}
\end{equation}
Using the Minkowski metric~\eqref{metric}, the contravariant temporal and spatial components of the velocity vector~\eqref{temporal velocity contravariant} and~\eqref{spatial velocity contravariant and covariant}, and the partial time derivative~\eqref{temporal partial derivative contravariant}, the relativistic momentum continuity equation~\eqref{relativistic momentum continuity equation fluid i 0} is recast as,
\begin{equation}\label{relativistic momentum continuity equation fluid i 1}
\partial_{t}\left(\gamma^2\left(\frac{\tilde{e} + \tilde{p}}{c^2}\right)v^{i}\right) + \partial_{j}\left(\gamma^2\left(\frac{\tilde{e} + \tilde{p}}{c^2}\right)v^{i}v^{j}\right) + \partial^{i}\tilde{p} = f^{i}
\end{equation}
According to the frame duality~\eqref{duality space}, the contraction of the relativistic momentum continuity equation~\eqref{relativistic momentum continuity equation fluid i 1} with the spatial orthonormal basis vectors $\boldsymbol{e}_i$ is written as,
\begin{equation}\label{relativistic momentum continuity equation fluid i 2}
\begin{split}
&\partial_{t}\left(\gamma^2\left(\frac{\tilde{e} + \tilde{p}}{c^2}\right)v^{i}\,\boldsymbol{e}_i\right) + \left(\boldsymbol{e}^k\,\partial_{k}\right)\cdot\left(\gamma^2\left(\frac{\tilde{e} + \tilde{p}}{c^2}\right)v^{i}v^{j}\,\boldsymbol{e}_i\,\otimes\,\boldsymbol{e}_j\right)\\
&+ \left(\boldsymbol{e}^i\,\partial_i\right)\tilde{p} = f^{i}\,\boldsymbol{e}_i
\end{split}
\end{equation}
since $\boldsymbol{e}^i\,\partial_{i} = \boldsymbol{e}_i\,\partial^{i}$. In view of the frame duality~\eqref{duality space}, the spatial velocity vector~\eqref{spatial velocity}, the gradient~\eqref{gradient}, the transformation laws for the energy density~\eqref{density e frames} and the pressure~\eqref{pressure frames}, and the force density spatial vector~\eqref{force density spatial}, the relativistic momentum continuity equation~\eqref{relativistic momentum continuity equation fluid i 2} is recast as,
\begin{equation}\label{relativistic momentum continuity equation fluid i 3}
\partial_{t}\left(\left(\frac{e + p}{c^2}\right)\,\boldsymbol{v}\right) + \boldsymbol{\nabla}\cdot\left(\left(\frac{e + p}{c^2}\right)\,\boldsymbol{v}\,\otimes\,\boldsymbol{v}\right) + \boldsymbol{\nabla}\,\tilde{p} = \boldsymbol{f}
\end{equation}
Using the decomposition~\eqref{internal energy density bis} of the energy density $e$, and the enthalpy density $h$ in the inertial frame $\mathcal{R}$,
\begin{equation}\label{enthalpy density}
h = u + p
\end{equation}
the relativistic momentum continuity equation~\eqref{relativistic momentum continuity equation fluid i 3} becomes,
\begin{equation}\label{relativistic momentum continuity equation fluid i 4}
\partial_{t}\left(\left(\rho + \frac{h}{c^2}\right)\boldsymbol{v}\right) + \boldsymbol{\nabla}\cdot\left(\left(\rho + \frac{h}{c^2}\right)\boldsymbol{v}\,\otimes\,\boldsymbol{v}\right) + \boldsymbol{\nabla}\,\tilde{p} = \boldsymbol{f}
\end{equation}
The symmetric spatial stress tensor in the inertial frame $\mathcal{R}$ is defined as,
\begin{equation}\label{spatial stress tensor 0}
\boldsymbol{\tau} = -\,\left(\rho + \frac{h}{c^2}\right)\left(\boldsymbol{v}\,\otimes\,\boldsymbol{v}\right) -\,\tilde{p}\,\mathbb{1}_3
\end{equation}
where $\mathbb{1}_3$ is the spatial identity rank-2 tensor. In view of the momentum density~\eqref{momentum density}, the spatial stress tensor~\eqref{spatial stress tensor 0} in the inertial frame $\mathcal{R}$ becomes,
\begin{equation}\label{spatial stress tensor}
\boxed{\boldsymbol{\tau} = -\,\boldsymbol{p}\,\otimes\,\boldsymbol{v} -\,\tilde{p}\,\mathbb{1}_3 -\,h\left(\frac{\boldsymbol{v}}{c}\,\otimes\,\frac{\boldsymbol{v}}{c}\right)  \qquad\textrm{(relativistic)}}
\end{equation}
where the last term is a relativistic correction of the spatial stress tensor which leads to a relativistic coupling between heat and work. In view of the momentum density vector~\eqref{momentum density} and the spatial stress tensor~\eqref{spatial stress tensor}, the relativistic momentum continuity equation~\eqref{relativistic momentum continuity equation fluid i 4} in the inertial frame $\mathcal{R}$ is recast as,
\begin{equation}\label{relativistic spatial momentum continuity equation fluid}
\boxed{\partial_t\left(\boldsymbol{p} + \frac{h}{c^2}\,\boldsymbol{v}\right) -\,\boldsymbol{\nabla}\cdot\boldsymbol{\tau} = \boldsymbol{f} \qquad\textrm{(relativistic)}}
\end{equation}
where the relativistic correction to the momentum density $\boldsymbol{p}$ also leads to a relativistic coupling between heat and work. In the non-relativistic limit, the momentum continuity equation~\eqref{relativistic spatial momentum continuity equation fluid} reduces to,~\cite{Brechet:2019}
\begin{equation}\label{non relativistic spatial momentum continuity equation}
\boxed{\partial_t\,\boldsymbol{p} -\,\boldsymbol{\nabla}\cdot\boldsymbol{\tau} = \boldsymbol{f} \qquad\textrm{(non-relativistic)}}
\end{equation}
and the spatial stress tensor~\eqref{spatial stress tensor} reduces to,
\begin{equation}\label{non relativistic spatial stress tensor}
\boxed{\boldsymbol{\tau} = -\,\boldsymbol{p}\,\otimes\,\boldsymbol{v} -\,p\,\mathbb{1}_3 \qquad\textrm{(non-relativistic)}}
\end{equation}
as expected.
%


\section{Relativistic inertia continuity equation}
\label{Relativistic inertia continuity equation}

\noindent The entropy~\eqref{entropy S} and the number of moles of substance~\eqref{number N} are frame independent scalar quantities. By contrast, the mass~\eqref{mass M frames} is a frame dependent quantity. Thus, the mass continuity equation cannot be established by following the same theoretical approach as for the entropy in \S~\ref{Relativistic second law of thermodynamics} and for the quantity of substance in \S~\ref{Relativistic matter continuity equation}. In fact, since the enthalpy density $h$ of the fluid carries inertia, the matter continuity equation is established in this section for the relativistic inertia density $m = \rho + h/c^2$. This is achieved by projecting the relativistic energy-momentum continuity equation~\eqref{relativistic momentum continuity equation} along the worldline of the local element of fluid defined by the velocity in the inertial frame $\mathcal{R}$. The contraction of the relativistic energy-momentum continuity equation~\eqref{relativistic momentum continuity equation} with the velocity vector reads,
\begin{equation}\label{relativistic momentum continuity equation contracted}
\left(\partial_{\mu}\,T^{\mu\nu}\right)u_{\nu} = f^{\nu}\,u_{\nu}
\end{equation}
In view of the orthogonality condition~\eqref{velocity force} between the force and velocity vectors and the contravariant components of the force vector~\eqref{force density contravariant}, we obtain the identity,
\begin{equation}\label{power density identity}
F^{\nu}\,u_{\nu} = \left(f^{\nu}\,u_{\nu}\right)d\tilde{V}_3 = 0
\end{equation}
According to the identity~\eqref{power density identity}, the contracted relativistic energy-momentum continuity equation~\eqref{relativistic momentum continuity equation contracted} reduces to,
\begin{equation}\label{relativistic momentum continuity equation contracted 2}
u_{\nu}\left(\partial_{\mu}\,T^{\mu\nu}\right) = 0
\end{equation}
In view of the contravariant components of the stress energy momentum tensor~\eqref{stress-energy-momentum tensor fluid mu nu}, the contracted relativistic energy-momentum continuity equation~\eqref{relativistic momentum continuity equation contracted 2} becomes,
\begin{equation}\label{relativistic momentum continuity equation contracted 3}
u_{\nu}\,\partial_{\mu}\left(\tilde{m}\,u^{\mu}\,u^{\nu}\right) + u_{\nu}\,\partial_{\mu}\left(\eta^{\mu\nu}\,\tilde{p}\right) = 0
\end{equation}
In view of the contraction of the velocities~\eqref{velocities contraction}, the relativistic contracted energy-momentum  continuity equation~\eqref{relativistic momentum continuity equation contracted 3} is recast as,
\begin{equation}\label{relativistic momentum continuity equation contracted 4}
-\,c^2\partial_{\mu}\left(\tilde{m}\,u^{\mu}\right) + \tilde{m}\,u^{\mu}\left(u_{\nu}\,\partial_{\mu}\,u^{\nu}\right) + u^{\mu}\,\partial_{\mu}\,\tilde{p} = 0
\end{equation}
According to the contraction of the velocities~\eqref{velocities contraction}, we obtain the identity,
\begin{equation}\label{derivative velocity contraction}
\partial_{\mu}\left(u_{\nu}\,u^{\nu}\right) = -\,\partial_{\mu}\left(c^2\right) = 2\,u_{\nu}\,\partial_{\mu}\,u^{\nu} = 0
\end{equation}
In view of the identity~\eqref{derivative velocity contraction}, the contracted relativistic energy-momentum continuity equation~\eqref{relativistic momentum continuity equation contracted 4} yields the relativistic inertia continuity equation,
\begin{equation}\label{relativistic inertia continuity equation}
\partial_{\mu}\left(\tilde{m}\,u^{\mu}\right) = u^{\mu}\,\partial_{\mu}\left(\frac{\tilde{p}}{c^2}\right)
\end{equation}
According to the inertia density~\eqref{inertia density} and the enthalpy density in the local rest frame $\mathcal{\tilde{R}}$,
\begin{equation}\label{enthalpy density rest}
\tilde{h} = \tilde{u} + \tilde{p}
\end{equation}
the relativistic inertia continuity equation~\eqref{relativistic inertia continuity equation} is recast as,
\begin{equation}\label{relativistic inertia continuity equation rho}
\boxed{\partial_{\mu}\left(\tilde{\rho}\,u^{\mu}\right) = -\,\partial_{\mu}\left(\frac{\tilde{h}}{c^2}\,u^{\mu}\right) + u^{\mu}\,\partial_{\mu}\left(\frac{\tilde{p}}{c^2}\right)\qquad\textrm{(relativistic)}}
\end{equation}
The relativistic inertia continuity equation~\eqref{relativistic inertia continuity equation rho} differs from the relativistic mass continuity equation, i.e. $\partial_{\mu}\left(\tilde{\rho}\,u^{\mu}\right) = 0$, that is usually assumed to hold in textbooks on relativity.~\cite{Hobson:2006,Schutz:2009} As shown here, such an assumption is not warranted since the right hand side of the relativistic inertia continuity equation~\eqref{relativistic inertia continuity equation rho} clearly does not vanish. Since the Lorentz factor $\gamma$ is a first-order relativistic correction factor, the derivative of the Lorentz factor $\partial_\mu\,\gamma$ is a second-order relativistic correction to the dynamics that can be neglected to first-order. Thus, in view of the relativistic transformation law for the pressure~\eqref{pressure frames}, the relativistic inertia continuity equation~\eqref{relativistic inertia continuity equation rho} multiplied by $\gamma$ can be recast as,
\begin{equation}\label{relativistic inertia continuity equation bis}
\partial_{\mu}\left(\gamma\,\tilde{m}\,u^{\mu}\right) = \gamma\,u^{\mu}\,\partial_{\mu}\left(\frac{\tilde{p}}{c^2}\right) = \frac{1}{\gamma}\,u^{\mu}\,\partial_{\mu}\left(\frac{p}{c^2}\right)
\end{equation}
In view of the frame duality~\eqref{duality space}, the contravariant time component of the velocity vector~\eqref{temporal velocity contravariant}, the spatial velocity vector~\eqref{spatial velocity}, the contravariant spatial components of the velocity vector~\eqref{spatial velocity contravariant and covariant}, the partial time derivative~\eqref{temporal partial derivative contravariant} and the gradient~\eqref{gradient}, we obtain the identities,
\begin{align}
\label{mathematical mass density}
&\partial_{\mu}\left(\gamma\,\tilde{m}\,u^{\mu}\right) = \partial_{t}\left(\gamma^2\tilde{m}\right) + \left(\partial_j\,\boldsymbol{e}^{j}\right)\cdot\left(\gamma^2\tilde{m}\,v^k\,\boldsymbol{e}_{k}\right) = \partial_{t}\,m + \boldsymbol{\nabla}\cdot\left(m\,\boldsymbol{v}\right)\\
\label{dot p}
&u^{\mu}\,\partial_{\mu}\left(\frac{p}{c^2}\right) = \gamma\left(\partial_t + \boldsymbol{v}\cdot\boldsymbol{\nabla}\right)\frac{p}{c^2}
\end{align}
The total time derivative of the pressure in the inertial frame $\mathcal{R}$ is defined as,
\begin{equation}\label{dot p bis}
\dot{p} = \left(\partial_t + \boldsymbol{v}\cdot\boldsymbol{\nabla}\right)p
\end{equation}
In view of the identities~\eqref{mathematical mass density} and~\eqref{dot p}, and the total time derivative of the pressure~\eqref{dot p bis}, the relativistic inertia continuity equation~\eqref{relativistic inertia continuity equation bis} in the inertial frame $\mathcal{\tilde{R}}$ is recast as,
\begin{equation}\label{relativistic inertia continuity equation inertial}
\boxed{\partial_{t}\,m + \boldsymbol{\nabla}\cdot\left(m\,\boldsymbol{v}\right) = \frac{\dot{p}}{c^2} \qquad\textrm{(relativistic)}}
\end{equation}
In the non-relativistic limit, the inertia continuity equation~\eqref{relativistic inertia continuity equation inertial} reduces to the mass continuity equation,~\cite{Brechet:2019}
\begin{equation}\label{non relativistic mass continuity equation inertial}
\boxed{\partial_{t}\,\rho + \boldsymbol{\nabla}\cdot\left(\rho\,\boldsymbol{v}\right) = 0 \qquad\textrm{(non-relativistic)}}
\end{equation}
as expected.
%


\section{Relativistic Euler equation}
\label{Relativistic Euler equation}

\noindent \noindent We deduce the relativistic Euler equation from the relativistic momentum continuity equation using the inertia density continuity equation in this section. In view of the inertia density~\eqref{inertia density} in the local rest frame $\mathcal{\tilde{R}}$, the inertia density in the inertial frame $\mathcal{R}$ is given by,
\begin{equation}\label{inertia density inertial}
m = \frac{e + p}{c^2}
\end{equation}
Using the transformation laws for the energy density~\eqref{density e frames} and the pressure~\eqref{pressure frames}, the inertia density~\eqref{inertia density inertial} is recast as,
\begin{equation}\label{inertia density inertial bis}
m = \gamma^2\left(\frac{\tilde{e} + \tilde{p}}{c^2}\right)
\end{equation}
The inertia density~\eqref{inertia density inertial bis} in the inertial frame $\mathcal{R}$ is related to the inertia density~\eqref{inertia density} in the local rest frame $\mathcal{\tilde{R}}$ by,
\begin{equation}\label{density m frames}
\boxed{m = \gamma^2\,\tilde{m}}
\end{equation}
In view of the inertia density~\eqref{inertia density inertial}, the relativistic momentum continuity equation~\eqref{relativistic momentum continuity equation fluid i 3} reduces to,  
\begin{equation}\label{relativistic momentum continuity equation fluid i m}
\partial_{t}\left(m\,\boldsymbol{v}\right) + \boldsymbol{\nabla}\cdot\left(m\,\boldsymbol{v}\,\otimes\,\boldsymbol{v}\right) + \boldsymbol{\nabla}\,\tilde{p} = \boldsymbol{f}
\end{equation}
The relativistic momentum continuity equation~\eqref{relativistic momentum continuity equation fluid i m} is expanded as,
\begin{equation}\label{relativistic momentum continuity equation fluid expanded}
m\left(\partial_t\,\boldsymbol{v} + \boldsymbol{v}\cdot\boldsymbol{\nabla}\,\boldsymbol{v}\right) + \Big(\partial_t\,m + \boldsymbol{\nabla}\cdot\left(m\,\boldsymbol{v}\right)\Big)\,\boldsymbol{v} = \boldsymbol{f} -\,\boldsymbol{\nabla}\,\tilde{p}
\end{equation}
The spatial acceleration vector of a fluid is written as,
\begin{equation}\label{spatial acceleration bis}
\boldsymbol{a} = \boldsymbol{\dot{v}} = \left(\partial_t + \boldsymbol{v}\cdot\boldsymbol{\nabla}\right)\boldsymbol{v}
\end{equation}
According to the spatial acceleration vector~\eqref{spatial acceleration bis}, the relativistic momentum continuity equation~\eqref{relativistic momentum continuity equation fluid expanded} yields the relativistic Euler equation in the inertial frame $\mathcal{R}$,
\begin{equation}\label{relativistic Euler equation 0}
m\,\boldsymbol{a} + \Big(\partial_t\,m + \boldsymbol{\nabla}\cdot\left(m\,\boldsymbol{v}\right)\Big)\,\boldsymbol{v} = \boldsymbol{f} -\,\boldsymbol{\nabla}\,\tilde{p} 
\end{equation}
In view of the inertia density~\eqref{inertia density}, the enthalpy density~\eqref{enthalpy density} and the inertia continuity equation~\eqref{relativistic inertia continuity equation inertial}, the relativistic Euler equation~\eqref{relativistic Euler equation 0} in the inertial frame $\mathcal{R}$ is recast as,
\begin{equation}\label{relativistic Euler equation}
\boxed{\left(\rho + \frac{h}{c^2}\right)\,\boldsymbol{a} = \boldsymbol{f} -\,\boldsymbol{\nabla}\,\tilde{p} -\,\frac{\dot{p}}{c^2}\,\boldsymbol{v} \qquad\textrm{(relativistic)}}
\end{equation}
where the relativistic corrections are the terms that are functions of the enthalpy density $h$, which leads to a relativistic coupling between heat and work, or of the time derivative of the pressure $\dot{p}$. In the non-relativistic limit, the Euler equation~\eqref{relativistic Euler equation} reduces to,~\cite{Brechet:2019}
\begin{equation}\label{non relativistic Euler equation}
\boxed{\rho\,\boldsymbol{a} = \boldsymbol{f} -\,\boldsymbol{\nabla}\,p \qquad\textrm{(non-relativistic)}}
\end{equation}
as expected.
%


\section{Conclusion}
\label{Conclusion}

\noindent Using the covariant approach based on the divergence theorem in Minkowski space outlined by Stückelberg~\cite{Stueckelberg:2013}, we derived the relativistic continuity equations for Lorentz invariant scalar functions such as the entropy $S$ and the number of moles of substance $N$, and for the vector components $P^{\nu}$ of the energy momentum vector. This approach lead naturally to a relativistic formulation of the first law~\eqref{variation function P divergence} and second law~\eqref{variation function S divergence} of thermodynamics,
\begin{equation}\label{first and second laws of thermodynamics}
\Delta P^{\nu}_{i\rightarrow f} = F^{\nu}\,\Delta t_{i\rightarrow f} \qquad\textrm{and}\qquad \Delta S_{i\rightarrow f} = \Sigma_{S\,i\rightarrow f} \geqslant 0
\end{equation}
where the variation of the state functions $\Delta P^{\nu}_{i\rightarrow f}$ and $\Delta S_{i\rightarrow f}$ are due to the source terms $F^{\nu}\,\Delta t_{i\rightarrow f}$ and $\Sigma_{S\,i\rightarrow f}$ in an arbitrary inertial frame $\mathcal{R}$. On one hand, the entropy $S$ and the number of moles of substance $N$ are frame independent quantities. It implies that the relativistic continuity equations for these quantities~\eqref{relativistic entropy continuity equation bis} and~\eqref{relativistic matter continuity equation bis} are Lorentz-invariant. On the other hand, the energy-momentum vector is not frame independent. Thus, for a perfect fluid the relativistic continuity equations for these quantities~\eqref{relativistic energy continuity equation fluid} and~\eqref{relativistic spatial momentum continuity equation fluid} are not Lorentz-invariant,
\begin{equation}\label{relativistic continuity equations energy and momentum}
\begin{split}
&\partial_{t}\left(e + \frac{v^2}{c^2}\,p\right) + \boldsymbol{\nabla}\cdot\left(e\,\boldsymbol{v}\right) = \left(\boldsymbol{f} -\,\boldsymbol{\nabla}\,p\right)\cdot\boldsymbol{v} \qquad\textrm{(energy)}\\
&\partial_t\left(\boldsymbol{p} + \frac{h}{c^2}\,\boldsymbol{v}\right) -\,\boldsymbol{\nabla}\cdot\boldsymbol{\tau} = \boldsymbol{f} \qquad\textrm{(momentum)}
\end{split}
\end{equation}
which leads to relativistic corrections due to the relativistic motion at velocity $\boldsymbol{v}$ of the local rest frame $\mathcal{\tilde{R}}$ with respect to the inertial frame $\mathcal{R}$. These relativistic corrections couple the relativistic work and heat. The first relativistic correction is an additional power density term $\partial_{t}\left(\frac{v^2}{c^2}\,p\right)$ due to the time derivative of the pressure $p$ in the relativistic continuity equation for the energy. The second relativistic correction is an additional spatial force density term $\partial_t\left(\frac{h}{c^2}\,\boldsymbol{v}\right)$ due to the time derivative of the enthalpy density $h$ in the relativistic continuity equation for the momentum. These relativistic continuity equations are derived from the stress energy-momentum tensor~\eqref{stress-energy-momentum tensor fluid mu nu} that is written in contravariant components in the inertial frame $\mathcal{R}$ as,
\begin{equation}\label{stress-energy-momentum tensor fluid contrvariant}
T^{\mu\nu} = \tilde{m}\,u^{\mu}\,u^{\nu} + \tilde{p}\,\eta^{\mu\nu} = \left(\tilde{\rho} + \frac{1}{c^2}\left(\tilde{u} + \tilde{p}\right)\right)u^{\mu}\,u^{\nu} + \tilde{p}\,\eta^{\mu\nu}
\end{equation}
where the relativistic inertia density $\tilde{m}$ is due mainly to the mass density $\tilde{\rho}$ and also to the internal energy density $\tilde{u}$ and to the pressure $\tilde{p}$ that carry inertia for a relativistic motion. The difference between the inertia density $\tilde{m}$ and the mass density $\tilde{\rho}$ is a relativistic correction. Since the mass density $\rho$ is part of a component of a rank-two tensor, the relativistic continuity equation for the mass was deduced from the relativistic continuity equation for the energy momentum by projection along the worldline defined by the velocity vector. In fact, it is the relativistic continuity equation~\eqref{relativistic inertia continuity equation inertial} for the inertia that was derived in this manner,
\begin{equation}\label{relativistic continuity equation inertia}
\boxed{\partial_{t}\,m + \boldsymbol{\nabla}\cdot\left(m\,\boldsymbol{v}\right) = \frac{\dot{p}}{c^2} \qquad\textrm{(inertia)}}
\end{equation}
where the relativistic corrections are the difference between the inertia and mass densities, i.e. $m-\,\rho = \frac{h}{c^2}$, and the mass flow rate $\frac{\dot{p}}{c^2}$ due to the proper time derivative of the pressure $p$. In view of this relativistic continuity equation, we determined the relativistic Euler equation~\eqref{relativistic Euler equation},
\begin{equation}\label{relativistic Euler equation conclusion}
\left(\rho + \frac{h}{c^2}\right)\,\boldsymbol{a} = \boldsymbol{f} -\,\boldsymbol{\nabla}\,\tilde{p} -\,\frac{\dot{p}}{c^2}\,\boldsymbol{v} \qquad\textrm{(Euler equation)}
\end{equation}
The first relativistic correction is an additional mass density term $\frac{h}{c^2}$ due to the enthalpy density of the perfect fluid that carries inertia. The second relativistic correction is an additional spatial force density term $-\,\frac{\dot{p}}{c^2}\,\boldsymbol{v}$ due to the proper time derivative of the pressure $p$.\\

\noindent In our theoretical derivation of relativistic thermodynamics, we reached the conclusion that the relativistic transformation laws for the intensive quantities, namely the temperature~\eqref{temperature frames}, the pressure~\eqref{pressure frames} and the chemical potential~\eqref{chemical potential frames} are given by,
\begin{equation}\label{intensive frames}
T = \gamma\,\tilde{T} \qquad\textrm{and}\qquad p = \gamma^2\,\tilde{p} \qquad\textrm{and}\qquad \mu = \gamma\,\tilde{\mu}
\end{equation}
The relativistic transformation law for the temperature is in agreement with Ott~\cite{Ott:1963}, Einstein~\cite{Liu:1992} in his latter years and Arzelies~\cite{Arzelies:1965}. The relativistic transformation law for the pressure agrees with the result obtained by Sutcliffe~\cite{Sutcliffe:1965}. These laws for the temperature and the pressure are in agreement with the ``Gedankenexperiment'' where a black body moving at relativistic velocity emits photons orthogonally to the direction of motion. The relativitic transverse Doppler effect accounts for the frequency shift affecting the temperature and the radiation pressure. Moreover, in view of the relativistic transformation law for the volume~\eqref{volume contraction},
\begin{equation}\label{volume frames}
V = \gamma^{-1}\,\tilde{V}
\end{equation}
the relativistic transformation laws for the densities of extensive quantities, namely the entropy density~\eqref{density s frames}, the number density~\eqref{density n frames},
\begin{equation}\label{extensive entropy number}
s = \gamma\,\tilde{s} \qquad\quad\textrm{and}\qquad n = \gamma\,\tilde{n}
\end{equation}
in agreement with Einstein~\cite{Einstein:1907} and Planck~\cite{Planck:1908}. The relativistic transformation laws for the other densities of extensive quantities, namely the mass density~\eqref{density rho frames}, the energy density~\eqref{density e frames} and the internal energy density~\eqref{density u frames} are given by,
\begin{equation}\label{extensive frames}
\rho = \gamma^2\,\tilde{\rho} \qquad\textrm{and}\qquad e = \gamma^2\,\tilde{e} \qquad\textrm{and}\qquad u = \gamma^2\,\tilde{u}
\end{equation}
The difference in the behaviour of the relativistic transformation laws~\eqref{extensive entropy number} and~\eqref{extensive frames}, namely the power of the Lorentz factor $\gamma$, is due to the fact that the entropy density $s$ and the number density $n$ are components of vectors in Minkowski space whereas the mass density $\rho$, the energy density $e$ and the internal energy density $u$ are parts of a component of a rank-$2$ tensor in Minkowski space. Another important aspect of our derivation is the Lorentz invariance of the entropy source density~\eqref{entropy source density frame independent}, which is a consequence of the Lorentz invariance of the relativistic continuity equation for the entropy,
\begin{equation}\label{entropy frames invariance}
\rho_s = \tilde{\rho}_{\tilde{s}}
\end{equation}
It implies that the relativistic irreversible thermodynamics of a perfect fluid can be described in a frame independent manner. This opens interesting perspectives for the derivation of the relativistic transformation laws for physical laws such as Fourier's law~\cite{Li:2016} and Ohm's law~\cite{Starke:2016}, and physical effects like the Seebeck effect and the Nernst effect.\\

\noindent In view of the relativistic transformation laws~\eqref{intensive frames},~\eqref{extensive entropy number} and~\eqref{extensive frames}, the relativistic thermodynamics of a perfect fluid appears like a $\gamma$ game. Are you interested in playing the game ?


\section*{Acknowledgements}

\noindent The authors would like to thank Aude Maier for initiating this work as well as Jean-Philippe Ansermet and Fran\c{c}ois A. Reuse for insightful discussions.


\pagebreak

\appendix



\section{Relativistic kinematics}
\label{Relativistic kinematics}

\noindent We briefly recall in this appendix the basic notions of relativistic kinematics that are essential to build a consistent theory of relativistic thermodynamics. According to the fundamental postulate of special relativity, the speed of light $c$ is the same with respect to any inertial frame. To describe the underlying structure of space and time, we consider a spherical light wave emitted at time $t = t^{\prime} = 0$ in an inertial frame $\mathcal{R}$ and in another inertial frame $\mathcal{R}^{\prime}$ respectively. In Cartesian coordinates, the infinitesimal distances squared travelled by the light wave in the two inertial frames satisfies the condition,~\cite{Hobson:2006,Schutz:2009}
\begin{equation}\label{inf distance squared}
dx^2 + dy^2 + dz^2 -\,c^2dt^2 = dx^{\prime 2} + dy^{\prime 2} + dz^{\prime 2} -\,c^2dt^{\prime 2} = 0
\end{equation}
In view of the condition~\eqref{inf distance squared}, the infinitesimal frame-independent space-time interval squared $ds^2$ is defined as,
\begin{equation}\label{inf line element}
ds^2 = -\,c^2dt^2 + dx^2 + dy^2 + dz^2 
\end{equation}
and called the infinitesimal line element. It vanishes for a light wave, i.e. $ds^2 = 0$, and it is negative for a point mass moving slower than light, i.e. $ds^2 < 0$. With the formal notation in contravariant space-time coordinates, 
\begin{equation}\label{contravariant notation}
dx^0 = c\,dt\,, \qquad dx^1 = dx\,, \qquad dx^2 = dy\,, \qquad\textrm{and}\qquad dx^3 = dz
\end{equation}
the infinitesimal line element~\eqref{inf line element} is recast as,
\begin{equation}\label{inf line element bis}
ds^2 = -\,\left(dx^{0}\right)^2 + \left(dx^{1}\right)^2 + \left(dx^{2}\right)^2 + \left(dx^{3}\right)^2 
\end{equation}
Using the Einstein summation convention, the infinitesimal line element~\eqref{inf line element bis} is written in contravariant coordinates $\{x^{\rho}\}$ in $\mathbb{R}^{1,3}$ as,~\cite{Weinberg:1972}
\begin{equation}\label{line element}
ds^2 = \eta_{\mu\nu}\,dx^{\mu}\,dx^{\nu}
\end{equation}
where $\mu,\nu \in \{0,1,2,3\}$ and the components of the Minkowski flat metric read,
\begin{equation}\label{metric}
\eta_{\mu\nu} = \delta_{\mu\nu} -\,2\,\delta_{\mu 0}\,\delta_{\nu 0}
\end{equation}
which defines the space-time signature $\left(-,+,+,+\right)$. The set of orthonormal basis vectors in flat space-time $\mathbb{R}^{1,3}$ is $\{\boldsymbol{e}_{\mu}\}$ where $\mu\in\{0,1,2,3\}$ and the set of orthonormal dual basis vectors is $\{\boldsymbol{e}^{\nu}\}$ where $\nu\in\{0,1,2,3\}$. This duality is written as,
\begin{equation}\label{duality space-time}
\boldsymbol{e}^{\nu}\cdot\boldsymbol{e}_{\mu} = \delta^{\nu}_{\ \mu}
\end{equation}
The set of orthonormal basis vectors in space $\mathbb{R}^{3}$ is $\{\boldsymbol{e}_{j}\}$ where $j\in\{1,2,3\}$ and the set of orthonormal dual basis vectors is $\{\boldsymbol{e}^{k}\}$ where $k\in\{1,2,3\}$. This duality is written as,
\begin{equation}\label{duality space}
\boldsymbol{e}^{k}\cdot\boldsymbol{e}_{j} = \delta^{k}_{\ j}
\end{equation}
The metric is given by the scalar product of basis vectors in space-time and in Minkowski space,~\cite{Misner:1973}
\begin{equation}\label{metric scalar product}
\boldsymbol{e}_{\mu}\cdot\boldsymbol{e}_{\nu} = \eta_{\mu\nu} \qquad\textrm{and}\qquad \boldsymbol{e}_{j}\cdot\boldsymbol{e}_{k} = \delta_{jk}
\end{equation}
where $\delta_{jk} = \eta_{jk}$. The infinitesimal spatial displacement vector $d\boldsymbol{x}$ is expressed in contravariant components in the orthonormal vector basis frame or in covariant components in the orthonormal dual vector basis frame as,
\begin{equation}\label{spatial displacement}
d\boldsymbol{x} = dx^j\,\boldsymbol{e}_{j} = dx_k\,\boldsymbol{e}^{k}
\end{equation}
The infinitesimal temporal displacement is expressed in contravariant and covariant coordinates as,
\begin{equation}\label{temporal displacement}
dx^0 = c\,dt \qquad\textrm{and}\qquad dx_0 = \eta_{0\mu}\,dx^{\mu} = \eta_{00}\,dx^{0} = -\,c\,dt
\end{equation}
In view of the infinitesimal spatial displacement~\eqref{spatial displacement} and the infinitesimal temporal displacement~\eqref{temporal displacement}, the infinitesimal line element~\eqref{line element} is recast as,
\begin{equation}\label{line element bis}
ds^2 = dx_0\,dx^0 + dx_j\,dx^j = -\,\left(cdt\right)^2 + \left(dx\right)^2
\end{equation}
where $dx^2 = d\boldsymbol{x}\cdot d\boldsymbol{x}$. In the rest frame $\mathcal{\tilde{R}}$, time is the proper time $\tilde{t}$ and there is no spatial displacement, i.e. $\left(d\tilde{x}\right)^2 = 0$. Thus, the infinitesimal line element~\eqref{line element bis} reduces to,~\cite{Weinberg:1972}
\begin{equation}\label{line element rest}
ds^2 = -\,c^2\,d\tilde{t}^2
\end{equation}
In view of the infinitesimal line elements~\eqref{line element} and~\eqref{line element rest}, the time derivative of the line element is given by,
\begin{equation}\label{line element derivative}
\frac{ds^2}{d\tilde{t}^2} = \eta_{\mu\nu}\,\frac{dx^{\mu}}{d\tilde{t}}\,\frac{dx^{\nu}}{d\tilde{t}} = -\,c^2
\end{equation}
The contravariant components of the velocity vector of a particle are defined as,~\cite{Carroll:2019}
\begin{equation}\label{velocity contravariant}
u^{\mu} = \frac{dx^{\mu}}{d\tilde{t}}
\end{equation}
The covariant components of the velocity vector are given by,
\begin{equation}\label{velocity covariant}
u_{\mu} = \eta_{\mu\nu}\,\frac{dx^{\nu}}{d\tilde{t}}
\end{equation}
In view of the time derivative of the infinitesimal line element~\eqref{line element derivative}, the contravariant~\eqref{velocity contravariant} and covariant~\eqref{velocity covariant} components of the velocity vector, the contraction of these components is frame independent,
\begin{equation}\label{velocities contraction}
u_{\mu}\,u^{\mu} = -\,c^2
\end{equation}
The Lorentz factor $\gamma$ is defined as,~\cite{Hobson:2006,Schutz:2009}
\begin{equation}\label{Lorentz factor}
\gamma = \frac{dt}{d\tilde{t}} \qquad\textrm{thus}\qquad dt = \gamma\,d\tilde{t}
\end{equation}
which describes time dilatation. The contravariant components of the velocity vector~\eqref{velocity contravariant} can be recast in terms of the Lorentz factor~\eqref{Lorentz factor} as,
\begin{equation}\label{velocity contravariant bis}
u^{\mu} = \frac{dx^{\mu}}{dt}\,\frac{dt}{d\tilde{t}} = \gamma\,\frac{dx^{\mu}}{dt}
\end{equation}
The covariant components of the velocity vector~\eqref{velocity covariant} can also be recast in terms of the Lorentz factor~\eqref{Lorentz factor} as,
\begin{equation}\label{velocity covariant bis}
u_{\mu} = \eta_{\mu\nu}\,\frac{dx^{\nu}}{dt}\,\frac{dt}{d\tilde{t}} = \gamma\,\frac{dx_{\mu}}{dt}
\end{equation}
In view of relations of the contravariant components of the velocity vector~\eqref{velocity contravariant} and the Lorentz factor~\eqref{Lorentz factor}, the contravariant time component of the velocity vector is written as,
\begin{equation}\label{temporal velocity contravariant}
u^0 = \frac{dx^0}{d\tilde{t}} = c\,\frac{dt}{d\tilde{t}} = \gamma\,c
\end{equation}
and the covariant time component of the velocity vector is given by,
\begin{equation}\label{temporal velocity covariant}
u_0 = \eta_{0\mu}\,u^{\mu} = \eta_{00}\,u^{0} = -\,\gamma\,c
\end{equation}
The spatial velocity vector $\boldsymbol{v}$ is expressed in contravariant components in the orthonormal basis frame as,
\begin{equation}\label{spatial velocity}
\boldsymbol{v} = v^j\,\boldsymbol{e}_{j} = \frac{dx^j}{dt}\,\boldsymbol{e}_{j} = \frac{dx^j}{d\tilde{t}}\,\frac{d\tilde{t}}{dt}\,\boldsymbol{e}_{j} =  \frac{1}{\gamma}\,u^j\,\boldsymbol{e}_{j}
\end{equation}
or in covariant components in the orthonormal dual basis frame as $v^j = v_j\,\boldsymbol{e}^{j}$. According to the contravariant components of the velocity vector~\eqref{velocity contravariant bis} and the spatial velocity vector~\eqref{spatial velocity}, the contravariant and covariant component of the velocity vector are written as,
\begin{equation}\label{spatial velocity contravariant and covariant}
u^j = \gamma\,v^j \qquad\textrm{and}\qquad u_j = \gamma\,v_j
\end{equation}
In view of the frame duality~\eqref{duality space}, the contravariant~\eqref{temporal velocity contravariant} and covariant~\eqref{temporal velocity covariant} time components of the velocity and the spatial components~\eqref{spatial velocity contravariant and covariant} of the velocity, the contraction of the velocities~\eqref{velocities contraction} is recast as,
\begin{equation}\label{velocities contraction bis}
u_{\mu}\,u^{\mu} = u_0\,u^0 + u_{i}\,u^{i} = \gamma^2\,\Big(-\,c^2 + \left(v_{j}\,\boldsymbol{e}^{j}\right)\cdot\left(v^{k}\,\boldsymbol{e}_{k}\right)\Big) = \gamma^2\left(-\,c^2 + v^2\right)
\end{equation}
where $v^2 = \boldsymbol{v}\cdot\boldsymbol{v}$. The identification of the velocity contractions~\eqref{velocities contraction} and~\eqref{velocities contraction bis} yields a kinematic expression~\cite{Einstein:1905} for the Lorentz factor~\eqref{Lorentz factor},
\begin{equation}\label{Lorentz factor v}
\gamma \equiv \gamma\left(v\right) = \left(1-\,\frac{v^2}{c^2}\right)^{-\,1/2}
\end{equation}
The contravariant components of the acceleration vector are defined as,~\cite{Schutz:2009}
\begin{equation}\label{acceleration contravariant 0}
a^{\mu} = \frac{du^{\mu}}{d\tilde{t}} = \frac{dx^{\nu}}{d\tilde{t}}\,\frac{du^{\mu}}{dx^{\nu}} = u^{\nu}\,\partial_{\nu}\,u^{\mu}
\end{equation}
According to the velocity contraction~\eqref{velocities contraction} and the contravariant components of the acceleration vector~\eqref{acceleration contravariant 0}, the velocity and acceleration vectors are orthogonal, which is written in components as,
\begin{equation}\label{velocity acceleration}
u_{\mu}\,a^{\mu} = u_{\mu}\,u^{\nu}\,\partial_{\nu}\,u^{\mu} = \frac{1}{2}\,u^{\nu}\,\partial_{\nu}\left(u_{\mu}\,u^{\mu}\right) = -\,\frac{1}{2}\,u^{\nu}\,\partial_{\nu}\,c^2 = 0
\end{equation}
According to the Lorentz factor~\eqref{Lorentz factor}, the contravariant components of the acceleration vector are recast as,
\begin{equation}\label{acceleration contravariant}
a^{\mu} = \frac{du^{\mu}}{d\tilde{t}} = \frac{dt}{d\tilde{t}}\,\frac{du^{\mu}}{dt} = \gamma\,\frac{du^{\mu}}{dt}
\end{equation}
In an inertial frame $\mathcal{R}$, the infinitesimal hypervolume in space-time is expressed as,
\begin{equation}\label{infinitesimal volume}
dV_4 = dx^{0}\,dx^{1}\,dx^{2}\,dx^{3} = c\,dt\,dV_3
\end{equation}
where $dV_3$ is the infinitesimal spatial volume in this frame. In the rest frame $\mathcal{\tilde{R}}$, the infinitesimal hypervolume in space-time is expressed as,
\begin{equation}\label{infinitesimal volume rest}
d\tilde{V}_4 = d\tilde{x}^{0}\,d\tilde{x}^{1}\,d\tilde{x}^{2}\,d\tilde{x}^{3} = c\,d\tilde{t}\,d\tilde{V}_3
\end{equation}
where we denote physical quantities evaluated in the rest frame $\mathcal{\tilde{R}}$ with a tilde. The infinitesimal hypervolume in the rest frame is related to the infinitesimal hypervolume in the inertial frame $\mathcal{R}$ by,~\cite{Carroll:2019}
\begin{equation}\label{infinitesimal volumes transformation}
d\tilde{x}^{0}\,d\tilde{x}^{1}\,d\tilde{x}^{2}\,d\tilde{x}^{3} = \frac{\partial\left(\tilde{x}^{0},\,\tilde{x}^{1},\,\tilde{x}^{2},\,\tilde{x}^{3}\right)}{\partial\left(x^{0},\,x^{1},\,x^{2},\,x^{3}\right)}\ 
dx^{0}\,dx^{1}\,dx^{2}\,dx^{3}
\end{equation}
where the Jacobian is the determinant of the Lorentz transformation $\Lambda$,
\begin{equation}\label{Jacobian}
\frac{\partial\left(\tilde{x}^{0},\,\tilde{x}^{1},\,\tilde{x}^{2},\,\tilde{x}^{3}\right)}{\partial\left(x^{0},\,x^{1},\,x^{2},\,x^{3}\right)} = \det\left(\Lambda\right)
\end{equation}
and the components of the Lorentz transformation are given by,
\begin{equation}\label{Lorentz transformation}
\Lambda^{\mu}_{\phantom{\mu}\nu} = \frac{\partial \tilde{x}^{\mu}}{\partial x^{\nu}}
\end{equation}
For an orthochronous Lorentz transformation where the rest frame $\mathcal{\tilde{R}}$ is moving with respect to the inertial frame $\mathcal{R}$ at scalar velocity $v^i$ along the $i$-axis, the components are given by,~\cite{Weinberg:1972}
\begin{equation}\label{Lorentz transformation components}
\begin{split}
&\Lambda^{0}_{\phantom{0}0} = \gamma \qquad\textrm{and}\qquad \Lambda^{0}_{\phantom{0}j} = \Lambda^{j}_{\phantom{j}0} = -\,\frac{1}{c}\,u^j = -\,\frac{\gamma}{c}\,v^j = -\,\frac{\gamma}{c}\,v^i\,\delta^{j}_{\phantom{j}i}\\
&\Lambda^{j}_{\phantom{k}k} = \left(\gamma -\,1\right)\delta^{j}_{\phantom{j}i}\,\delta^{i}_{\phantom{i}k} + \delta^{j}_{\phantom{j}k}
\end{split}
\end{equation}
In view of the Lorentz factor~\eqref{Lorentz factor v} and the components of the Lorentz transformation~\eqref{Lorentz transformation components}, the determinant of the Lorentz transformation yields,
\begin{equation}\label{Lorentz transformation det}
\det\left(\Lambda\right) = \Lambda^{0}_{\phantom{0}0}\,\Lambda^{i}_{\phantom{i}i} -\,\Lambda^{0}_{\phantom{0}i}\,\Lambda^{i}_{\phantom{i}0} = \gamma^2\left(1-\,\frac{v^2}{c^2}\right) = 1
\end{equation}
where $v^2 = v_{j}\,v^{j}$. This means that a Lorentz transformation is an isometry in Minkowski space. The infinitesimal hypervolume in the inertial frame $\mathcal{R}$ is related to the infinitesimal hypervolume in the rest frame $\mathcal{\tilde{R}}$ by,~\cite{Carroll:2019}
\begin{equation}\label{infinitesimal volumes transformation inverse}
dx^{0}\,dx^{1}\,dx^{2}\,dx^{3} = \frac{\partial\left(x^{0},\,x^{1},\,x^{2},\,x^{3}\right)}{\partial\left(\tilde{x}^{0},\,\tilde{x}^{1},\,\tilde{x}^{2},\,\tilde{x}^{3}\right)}\ d\tilde{x}^{0}\,d\tilde{x}^{1}\,d\tilde{x}^{2}\,d\tilde{x}^{3}
\end{equation}
where the Jacobian is the determinant of the inverse orthochronous Lorentz transformation $\Lambda^{-1}$,
\begin{equation}\label{Jacobian inverse}
\frac{\partial\left(x^{0},\,x^{1},\,x^{2},\,x^{3}\right)}{\partial\left(\tilde{x}^{0},\,\tilde{x}^{1},\,\tilde{x}^{2},\,\tilde{x}^{3}\right)} = \left(\frac{\partial\left(\tilde{x}^{0},\,\tilde{x}^{1},\,\tilde{x}^{2},\,\tilde{x}^{3}\right)}{\partial\left(x^{0},\,x^{1},\,x^{2},\,x^{3}\right)}\right)^{-1} = \det\left(\Lambda\right)^{-1} = \det\left(\Lambda^{-1}\right)
\end{equation}
The inverse orthochronous Lorentz transformation is also an isometry in Minkowski space, i.e. $\det\left(\Lambda^{-1}\right) = 1$, in view of the determinant~\eqref{Lorentz transformation det} since,
\begin{equation}\label{det Jacobian inverse}
\det\left(\Lambda\right)\det\left(\Lambda^{-1}\right) = \det\left(\Lambda\,\Lambda^{-1}\right) = \det\left(\mathbb{1}_4\right) = 1
\end{equation}
For the orthochronous inverse Lorentz transformation where the inertial frame $\mathcal{R}$ is moving with respect to the rest frame $\mathcal{\tilde{R}}$ at velocity $-\,v^i$ along the $x^{i}$-axis, the components are given by,~\cite{Weinberg:1972}
\begin{equation}\label{Lorentz transformation inverse components}
\begin{split}
&\left(\Lambda^{-1}\right)^0_{\phantom{0}0} = \gamma \qquad\textrm{and}\qquad \left(\Lambda^{-1}\right)^{0}_{\phantom{0}j} = \left(\Lambda^{-1}\right)^{j}_{\phantom{j}0} = \frac{1}{c}\,u^j = \frac{\gamma}{c}\,v^j = \frac{\gamma}{c}\,v^i\,\delta^{j}_{\phantom{j}i}\\
&\left(\Lambda^{-1}\right)^{j}_{\phantom{k}k} = \left(\gamma -\,1\right)\delta^{j}_{\phantom{j}i}\,\delta^{i}_{\phantom{i}k} + \delta^{j}_{\phantom{j}k}
\end{split}
\end{equation}
According to the infinitesimal volumes in the inertial frame~\eqref{infinitesimal volume} and the rest frame~\eqref{infinitesimal volume rest}, the volume transformation law~\eqref{infinitesimal volumes transformation}, the Jacobian~\eqref{Jacobian} and the determinant of the Lorentz transformation~\eqref{Lorentz transformation det}, the infinitesimal hypervolume in space-time is invariant,
\begin{equation}\label{infinitesimal hypervolume}
dV_4 = d\tilde{V}_4
\end{equation}
In view of the infinitesimal volumes in the inertial frame~\eqref{infinitesimal volume} and the rest frame~\eqref{infinitesimal volume rest}, and the infinitesimal hypervolume in space-time~\eqref{infinitesimal hypervolume}, we obtain the identity,
\begin{equation}\label{infinitesimal hypervolume bis}
dt\,dV_3 = d\tilde{t}\,d\tilde{V}_3
\end{equation}
According to the time dilatation~\eqref{Lorentz factor} and the hypervolume identity~\eqref{infinitesimal hypervolume bis}, we obtain a volume contraction due to a length contraction in the direction of motion along the $i$-axis,~\cite{Hobson:2006}
\begin{equation}\label{volume contraction}
dV_3 = \frac{1}{\gamma}\,d\tilde{V}_3
\end{equation}
%


\section{Relativistic dynamics}
\label{Relativistic dynamics}

\noindent We briefly review in this appendix the foundations of the relativistic dynamics of a particle within a special relativistic framework. In particular, we derive the relativistic transformation laws for the mass and energy of the particle between the rest frame $\mathcal{\tilde{R}}$ and an inertial frame $\mathcal{R}$. The contravariant time component of the momentum vector of a particle in an inertial frame $\mathcal{R}$ is the energy up to a factor $c^{-1}$,~\cite{Hobson:2006}
\begin{equation}\label{temporal momentum contravariant}
P^0 = \frac{E}{c}
\end{equation}
and the covariant time component of the momentum vector in the inertial frame $\mathcal{R}$ is given by,
\begin{equation}\label{temporal momentum covariant}
P_0 = \eta_{0\mu}\,P^{\mu} = \eta_{00}\,P^{0} = -\,\frac{E}{c}
\end{equation}
The contravariant component of the momentum vector $P^{\mu}$ is the product of the mass $\tilde{M}$ of the particle in the rest frame $\mathcal{R}$ and the contravariant component of the velocity vector $u^{\mu}$,~\cite{Hartle:2003}
\begin{equation}\label{momentum contravariant}
P^{\mu} = \tilde{M}\,u^{\mu}
\end{equation}
In view of the contravariant time component of the velocity vector~\eqref{temporal velocity contravariant} and the contravariant component of the momentum vector~\eqref{momentum contravariant}, the energy for a relativistic particle~\eqref{temporal momentum covariant} is written as,~\cite{Hobson:2006}
\begin{equation}\label{energy}
E = c\,P^{0} = c\,\tilde{M}\,u^{0} = \gamma\,\tilde{M}\,c^2
\end{equation}
In view of the Lorentz factor~\eqref{Lorentz factor v}, the energy~\eqref{energy} is recast explicitly as,
\begin{equation}\label{energy bis}
E = \left(1 -\,\frac{v^2}{c^2}\right)^{-\,1/2}\tilde{M}\,c^2
\end{equation}
In the non-relativistic limit, to first-order in $v^2/c^2$, the energy~\eqref{energy bis} reduces to,
\begin{equation}\label{energy density non rel}
E \simeq \left(1 + \frac{1}{2}\,\frac{v^2}{c^2}\right)M\,c^2 = \tilde{M}\,c^2 + \frac{1}{2}\,\tilde{M}\,v^2 \quad\ \ \left(\textrm{non-relativistic}\right)
\end{equation}
as expected. The energy $\tilde{E}$ of the particle in the rest frame $\mathcal{\tilde{R}}$ is the rest energy,
\begin{equation}\label{energy rest}
\tilde{E} = \tilde{M}\,c^2
\end{equation}
The energy $E$ of the particle~\eqref{energy} in the inertial frame $\mathcal{R}$ is expressed in terms of the energy $\tilde{E}$ of the particle~\eqref{energy rest} in the rest frame $\mathcal{\tilde{R}}$ as,~\cite{Hobson:2006,Schutz:2009}
\begin{equation}\label{energy E frames}
\boxed{E = \gamma\,\tilde{E}}
\end{equation}
In view of the energy $\tilde{E}$ of the particle~\eqref{energy rest} in the rest frame $\mathcal{\tilde{R}}$, the energy $E$ of the particle in the inertial frame $\mathcal{R}$ is recast in terms of the mass $M$ in the inertial frame $\mathcal{R}$ as,
\begin{equation}\label{energy inertial}
E = M\,c^2
\end{equation}
According to the energy of the particle~\eqref{energy} and~\eqref{energy inertial}, the mass $M$ of the particle in the inertial frame $\mathcal{R}$ is expressed in terms of the mass $\tilde{M}$ of the particle in the rest frame $\mathcal{\tilde{R}}$ as,~\cite{Hobson:2006,Schutz:2009}
\begin{equation}\label{mass M frames}
\boxed{M = \gamma\,\tilde{M}}
\end{equation}
The spatial momentum vector $\boldsymbol{P}$ is expressed in contravariant components in the orthonormal basis frame or in covariant components in the orthonormal dual basis frame as,
\begin{equation}\label{momentum vector}
\boldsymbol{P} = P^j\,\boldsymbol{e}_{j} = P_k\,\boldsymbol{e}^{k}
\end{equation}
In view of the spatial velocity vector~\eqref{spatial velocity} and the contravariant components of the momentum vector~\eqref{momentum contravariant}, the spatial momentum vector~\eqref{momentum vector} is recast as,
\begin{equation}\label{momentum}
\boldsymbol{P} = P^j\,\boldsymbol{e}_{j} = \tilde{M}\,u^j\,\boldsymbol{e}_{j} = \gamma\,\tilde{M}\,v^j\,\boldsymbol{e}_{j} = \gamma\,\tilde{M}\,\boldsymbol{v}
\end{equation}
The covariant component of the momentum vector $P_{\mu}$ is the product of the mass $\tilde{M}$ in the rest frame $\mathcal{R}$ and the covariant component of the velocity vector $u_{\mu}$,
\begin{equation}\label{momentum covariant}
P_{\mu} = \tilde{M}\,u_{\mu}
\end{equation}
According to the contraction of the velocities~\eqref{velocities contraction}, the contravariant and covariant components of the momentum vector~\eqref{momentum contravariant} and~\eqref{momentum covariant}, the contraction of the momenta is expressed as,
\begin{equation}\label{Casimir invariant 1}
P_{\mu}\,P^{\mu} = \tilde{M}^2\,u_{\mu}\,u^{\mu} = -\,\tilde{M}^2c^2
\end{equation}
which is a Casimir invariant of the Poincaré algebra. In view of the frame duality~\eqref{duality space}, the contravariant and covariant temporal and spatial components of the momentum vector~\eqref{temporal momentum contravariant},~\eqref{temporal momentum covariant},~\eqref{momentum contravariant} and~\eqref{momentum covariant}, the Casimir invariant is recast as,
\begin{equation}\label{Casimir invariant 2}
P_{\mu}\,P^{\mu} = P_{0}\,P^{0} + P_{j}\,P^{j} = -\,\frac{E^2}{c^2} + \left(P_{j}\,\boldsymbol{e}^{j}\right)\cdot\left(P^{k}\,\boldsymbol{e}_{k}\right) = -\,\frac{E^2}{c^2} + P^2
\end{equation}
where $P^2 = \boldsymbol{P}\cdot\boldsymbol{P}$. The identification of the Casimir invariants~\eqref{Casimir invariant 1} and~\eqref{Casimir invariant 2} yields the energy-momentum invariant for the relativistic motion of a particle,~\cite{Hobson:2006}
\begin{equation}\label{energy momentum invariant}
E^2 = P^2c^2 + \tilde{M}^2c^4
\end{equation}
where $P^2c^2$ is the square of the relativistic kinetic energy and $\tilde{M}^2c^4$ is the square of the rest energy in the rest frame $\mathcal{\tilde{R}}$. The relativistic law of motion states that the force vector is the time derivative of the momentum vector. In view of the definition~\eqref{Lorentz factor} of the Lorentz factor, the contravariant components of the force vector are given by,~\cite{Hobson:2006,Weinberg:1972}
\begin{equation}\label{relativistic law of motion}
F^{\mu} = \frac{dP^{\mu}}{d\tilde{t}} = \frac{dP^{\mu}}{dt}\,\frac{dt}{d\tilde{t}} = \gamma\,\frac{dP^{\mu}}{dt}
\end{equation}
Since the mass $\tilde{M}$ in the rest frame $\mathcal{\tilde{R}}$ is constant, the contravariant components of the acceleration vector~\eqref{acceleration contravariant} and momentum vector~\eqref{momentum contravariant} imply that the contravariant components of the force vector can be recast as,
\begin{equation}\label{relativitic law of motion constant}
F^{\mu} = \frac{d}{d\tilde{t}}\left(\tilde{M}\,u^{\mu}\right) = \tilde{M}\,\frac{du^{\mu}}{d\tilde{t}} = \tilde{M}\,a^{\mu}
\end{equation}
In view of the orthogonality between the velocity and the acceleration vectors~\eqref{velocity acceleration} and the relativistic law of motion~\eqref{relativistic law of motion}, the velocity and force vectors are orthogonal,
\begin{equation}\label{velocity force}
u_{\mu}\,F^{\mu} = \tilde{M}\,u_{\mu}\,a^{\mu} = 0
\end{equation}
According to the non-relativistic law of motion, the spatial force vector $\boldsymbol{F}$ in the inertial frame $\mathcal{R}$ is expressed in contravariant components in the orthonormal basis frame as the time derivative of the spatial momentum vector~\eqref{momentum vector},
\begin{equation}\label{law of motion}
\boldsymbol{F} = \frac{d\boldsymbol{P}}{dt} = \frac{dP^j}{dt}\,\boldsymbol{e}_{j}
\end{equation}
In view of the contravariant components of the relativistic law of motion~\eqref{relativistic law of motion}, the spatial force vector in the inertial frame $\mathcal{R}$ can be recast as,
\begin{equation}\label{force}
\boldsymbol{F} = \frac{1}{\gamma}\,F^{j}\,\boldsymbol{e}_{j}
\end{equation}
In view of the contravariant time component of the momentum vector~\eqref{temporal momentum contravariant} and the relativistic law of motion~\eqref{relativistic law of motion}, the contravariant time component of the force vector is given by,
\begin{equation}\label{temporal force contravariant 0}
F^{0} = \gamma\,\frac{dP^{0}}{dt} = \frac{\gamma}{c}\,\frac{dE}{dt}
\end{equation}
According to the covariant time component of the velocity vector~\eqref{temporal velocity covariant}, the spatial components of the velocity vector~\eqref{spatial velocity contravariant and covariant} and the time component of the force vector~\eqref{temporal force contravariant 0}, the orthogonality condition between the velocity and force vectors~\eqref{velocity force} is written as,
\begin{equation}\label{velocity force bis}
u_{0}\,F^{0} + u_{j}\,F^{j} = -\,\gamma^2\,\frac{dE}{dt} + \gamma\,v_j\,F^j = 0 
\end{equation}
In view of the frame duality~\eqref{duality space}, the spatial velocity vector~\eqref{spatial velocity} and the force vector~\eqref{force}, the orthogonality condition between the velocity and force vectors~\eqref{velocity force bis} yields the time derivative of the energy,
\begin{equation}\label{time derivative energy}
\frac{dE}{dt} = \frac{1}{\gamma}\,F^j\,v_j = \left(\frac{1}{\gamma}\,F^j\,\boldsymbol{e}_j\right)\cdot\left(v_k\,\boldsymbol{e}^{k}\right) = \boldsymbol{F}\cdot\boldsymbol{v}
\end{equation}
According to the time derivative of the energy~\eqref{time derivative energy}, the contravariant time component of the force vector~\eqref{temporal force contravariant 0}, which represents the relativistic power exerted by the force, is recast as,~\cite{Hobson:2006,Weinberg:1972}
\begin{equation}\label{temporal force contravariant}
\boxed{F^{0} = \frac{\gamma}{c}\,\boldsymbol{F}\cdot\boldsymbol{v}}
\end{equation}
%


\bibliography{references}
\bibliographystyle{plainnat}

\end{document}